\documentclass[fleqn,usenatbib]{mnras}

\usepackage{newtxtext,newtxmath}

\usepackage[T1]{fontenc}

\DeclareRobustCommand{\VAN}[3]{#2}
\let\VANthebibliography\thebibliography
\def\thebibliography{\DeclareRobustCommand{\VAN}[3]{##3}\VANthebibliography}

\usepackage{graphicx}	% Including figure files
\usepackage{amsmath}	% Advanced maths commands
\usepackage{amssymb}	% Extra maths symbols
\usepackage{graphicx}
\usepackage{subcaption}
\usepackage{appendix}
\usepackage{multicol}
\usepackage{rotating}
\usepackage{pdflscape}

\title[NH$_2$D Cepheus Survey]{A Survey of Deuterated Ammonia in the Cepheus Star-Forming Region L1251}

\author[M. Galloway-Sprietsma et al.]{
Maria Galloway-Sprietsma,$^{1,2}$\thanks{E-mail: mgallowayspriets@ufl.edu (MGS)} %\newline Senior Honors Thesis at The University of Arizona}
Yancy L. Shirley,$^{2}$\thanks{E-mail: yshirley@email.arizona.edu (YLS)}
James Di Francesco$^{3}$\thanks{E-mail: james.difrancesco@nrc-cnrc.gc.ca}, Jared Keown$^{3}$\thanks{E-mail: jkeown@uvic.ca},
\newauthor
Samantha Scibelli$^{2}$\thanks{E-mail: sscibelli@email.arizona.edu}, Olli Sipil{\"a}$^{4}$\thanks{osipila@mpe.mpg.de}, Rachel Smullen$^{5}$\thanks{rsmullen@lanl.gov}
\\
$^{1}$Bryant Space Science Center, The University of Florida, 1772 Stadium Rd, Gainesville, FL 32611
\\
$^{2}$Steward Observatory, The University of Arizona, 933 N. Cherry Ave., Tucson, AZ 85721
\\
$^{3}$Herzberg Astronomy and Astrophysics Research Centre, National Research Council of Canada, 5071 West Saanich Road, Victoria, BC V9E 2E7 CANADA
\\
$^{4}$Max-Planck-Institut für extraterrestrische Physik, Giessenbachstrasse 1,
85748 Garching, Germany
\\
$^{5}$ Center for Theoretical Astrophysics, Los Alamos National Laboratory, Los Alamos, NM 87545, USA 
}

\pubyear{2021}

\begin{document}
\label{firstpage}
\pagerange{\pageref{firstpage}--\pageref{lastpage}}
\maketitle

\begin{abstract}
Understanding the chemical processes during starless core and prestellar core evolution is an important step in understanding the initial stages of star and disk formation. 
This project is a study of deuterated ammonia, o-NH$_2$D, in the L1251 star-forming region toward Cepheus. Twenty-two dense cores (twenty of which are starless or prestellar, and two of which have a protostar), previously identified by p-NH$_3$ (1,1) observations, were targeted  with the 12m Arizona Radio Observatory telescope on Kitt Peak.
o-NH$_2$D J$_{\rm{K_a} \rm{K_c}}^{\pm} =$ $1_{11}^{+} \rightarrow 1_{01}^{-}$ was detected in 13 (59\%) of the NH$_3$-detected cores with a median sensitivity of $\sigma_{T_{mb}} = 17$ mK.
All cores detected in o-NH$_2$D at this sensitivity have p-NH$_3$ column densities $> 10^{14}$ cm$^{-2}$.
The o-NH$_2$D column densities were calculated using the constant excitation temperature (CTEX) approximation while correcting for the filling fraction of the NH$_3$ source size.
The median deuterium fraction was found to be 0.11 (including 3$\sigma$ upper limits). However, there are no strong, discernible trends in plots of deuterium fraction with any physical or evolutionary variables.
If the cores in L1251 have similar initial chemical conditions, then this result is evidence of the cores physically evolving at different rates.
\end{abstract}


%%%%%%%%%%%%%%%%%%%%%%%%%%%%%%%%%%%%%%%%%%%%%%%%%%

%%%%%%%%%%%%%%%%% BODY OF PAPER %%%%%%%%%%%%%%%%%%

\section{Introduction}
Starless, prestellar, and protostellar cores are dense regions in molecular clouds with a rich molecular gas phase chemistry \citep{2007ARA&A..45..339B, 2007prpl.conf...17D, 2014prpl.conf...27A, 2014prpl.conf..195D,2020ARA&A..58..727J}.
Deuterated molecules allow us to observe cold and dense regions within cores to learn more about the chemical evolution of these cores early in the star and planet formation process. 
The chemistry during the starless and prestellar core phases have far-reaching effects for future disk and planet formation chemistry by setting the initial chemical conditions immediately prior to the formation of a disk and protostar (e.g., \citealt{2014Sci...345.1590C, 2021A&A...650A.172J}). 

Deuterated isotopologues of molecules, where one or more hydrogen atoms have been replaced by a deuterium atom, were first identified toward Orion with the detection of the DCN molecule \citep{1973ApJ...179L..57J}. 
They have been found in dense molecular cores over a wide range of densities and temperatures and are prevalent in cold ($T \sim 10$ K), dense ($n > 10^4$ cm$^{-3}$) regions \citep{2014prpl.conf..859C}. 
Fractionation of the molecular D/H ratios have been observed that can be many orders of magnitude larger (i.e. \citealt{2002ApJ...571L..55L, 2004A&A...416..159P}) than the local Galactic atomic D/H ratio of $1.5 \times 10^{-5}$ and primordial D/H ratio of $2.6 \times 10^{-5}$ (\citealt{2003SSRv..106...49L, 2007AJ....133.1625R, 2016ApJ...830..148C, 2016A&A...594A..13P}).
Gas-phase deuteration chemistry is primarily driven by ion-molecule chemistry where collisional ionization from cosmic rays and secondary electrons form H$_2^+$ which reacts with the most abundant molecule, H$_2$, to form the ion H$_3^+$ \citep{1974ApJ...188...35W, 2006PNAS..10312232K}.  At low temperatures, the forward reaction of the H$_3^+$ ion with HD is favored to form H$_2$D$^+$ via the exothermic reaction
H$_3^+$ + HD 
forms
%$\rightleftharpoons$ 
H$_2$D$^+$ + 240 \rm{K} (\citealt{1989ApJ...340..906M,2003ApJ...591L..41R}).
This ion can then react with other gas phase molecules to transfer D to those molecules or form new molecules (i.e. CO + H$_2$D$^+$ forms DCO$^+$; \citealt{2009ApJ...706L..52Y}).
Low temperatures, $T_k \sim 10 - 20$ K, in the dense regions ($> 10^4$ cm$^{-3}$) of the clouds also force the freeze-out of molecules like CO \citep{1998ApJ...507L.171W, 2002A&A...389L...6B,  2017ApJ...849...80C}.
Since CO is a significant destruction route for H$_2$D$^+$,
its depletion enhances the gas phase
H$_2$D$^+$ abundance.
Icy grain surface chemistry also plays a role in chemical deuteration \citep{1983A&A...119..177T}, and allows further formation of deuterated species 
(\citealt{Fedoseev_2014}). 
This gas phase and solid state chemistry at low temperatures leads to an increase in the overall molecular D/H ratios which then leads to the deuterium fractionation of other species.

NH$_3$ is a popular molecular tracer that has been used to probe the structure of dense cores \citep{1983ARA&A..21..239H, 1989ApJS...71...89B, 1999ApJS..125..161J, 2015ApJ...805..185S, 2017ApJ...843...63F} since its lowest inversion transition has an optically thin critical density of $2 \times 10^3$ cm$^{-3}$ at 10 K \citep{2015PASP..127..299S}.
It also resists depletion except in the coldest and densest regions of prestellar cores (\citealt{2009ApJ...697.1457F, 2011A&A...534A.122R, 2014ApJ...790..129C, 2022arXiv220501201P, 2022ApJ...929...13C}).
The singly-deuterated version of ammonia, NH$_2$D, is potentially an excellent probe of dense gas that has spent a significant amount of time at high densities and at low temperatures \citep{1985A&A...142L...1O, 2000A&A...356.1039T, 2000ApJ...535..227S,
2001ApJ...554..933S,
2003A&A...403L..25H,
2005A&A...438..585R, 2017A&A...600A..61H}.
It forms primarily in the gas phase from ion-molecule reactions (\citealt{2014A&A...562A..83L, 2018MNRAS.477.4454H}), such as NH$_3$ + H$_2$D$^+$ forms NH$_3$D$^+$, which then undergo dissociative recombination (DR) with a free electron to form NH$_2$D (see \citealt{2004JChPh.120.7391O} for the measurements of the DR branching ratios of NH$_4^+$; unfortunately, there are no published laboratory measurements of the branching ratio for DR of NH$_3$D$^+$).
NH$_2$D is an oblate  asymmetric top molecule with a permanent electric dipole moment mostly aligned with the c-axis ($\mu_a = -0.18$ D, $\mu_c = 1.463$ D; \citealt{1982JMoSp..93...83C}).  
Like NH$_3$, NH$_2$D inverts via quantum tunneling and this inversion splits rotational energy levels in the ground vibrational state into symmetric (v = $0s$) and anti-symmetric (v = $0a$) levels. 
The rotation-inversion energy levels are denoted by the quantum numbers J$_{\rm{K_a K_c}}^{\pm}$ where the $\pm$ superscript corresponds to the inversion symmetry substate ($+$ $\Rightarrow$ v = $0s$ and $-$ $\Rightarrow$ v = $0a$).
Furthermore, NH$_2$D has a mirror plane with C$_{\rm{s}}$ symmetry resulting in symmetric (ortho) and anti-symmetric (para) nuclear wavefunctions with respect to the interchange of the two symmetric hydrogen atoms. 
Due to Fermi-Dirac statistics, each rotation-inversion energy level is classified as either ortho (o-NH$_2$D) or para (p-NH$_2$D) depending on whether $(-1)^{K_a + i_s}$ is equal to $-1$ (ortho) or $+1$ (para) where $i_s = 0$ for v = $0s$ and $i_s = 1$ for v = $0a$.

\begin{figure*}
    \centering
    \hspace{-0.5cm}
    \includegraphics[width=56mm]{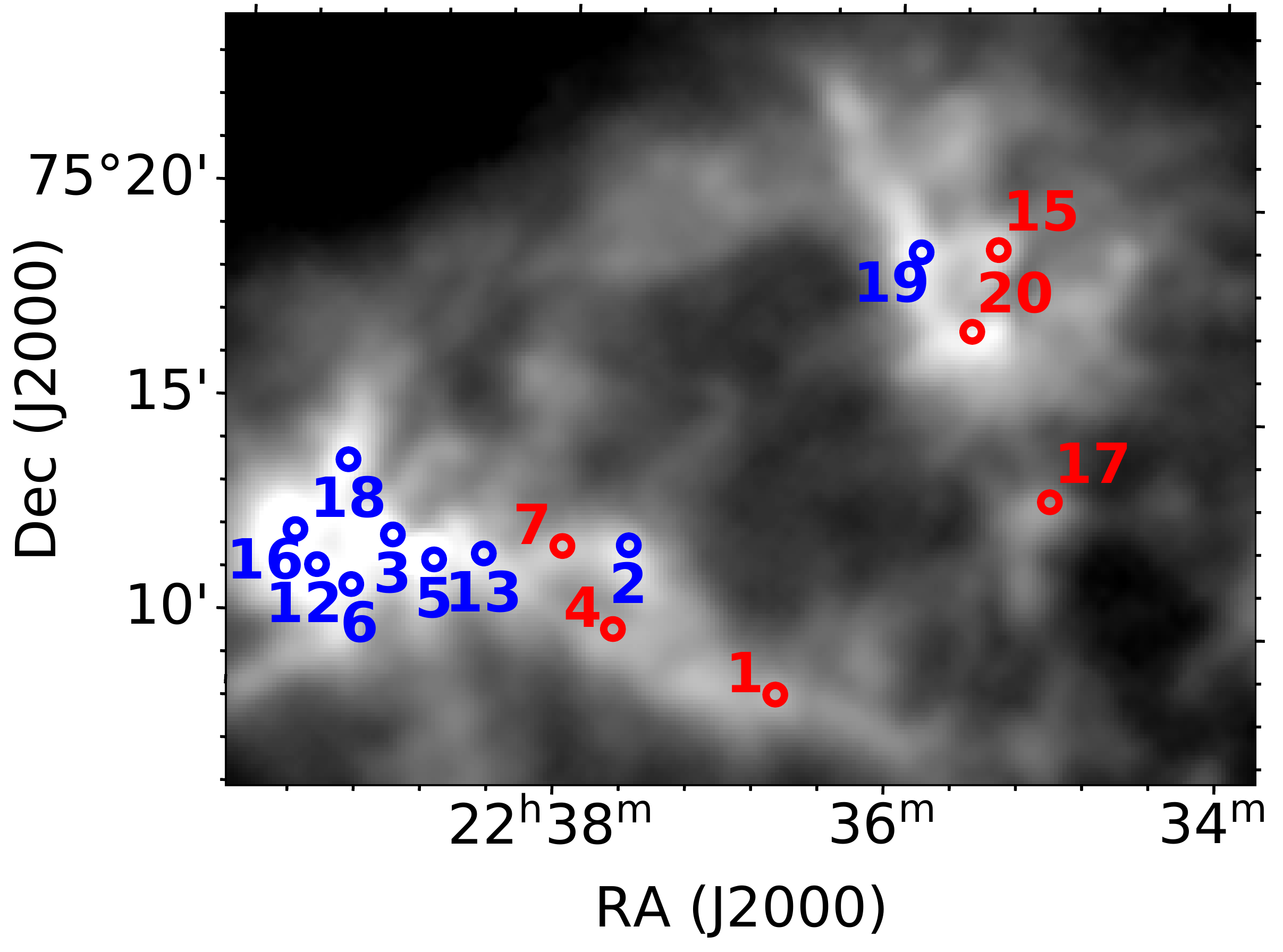}
    \includegraphics[width=56mm]{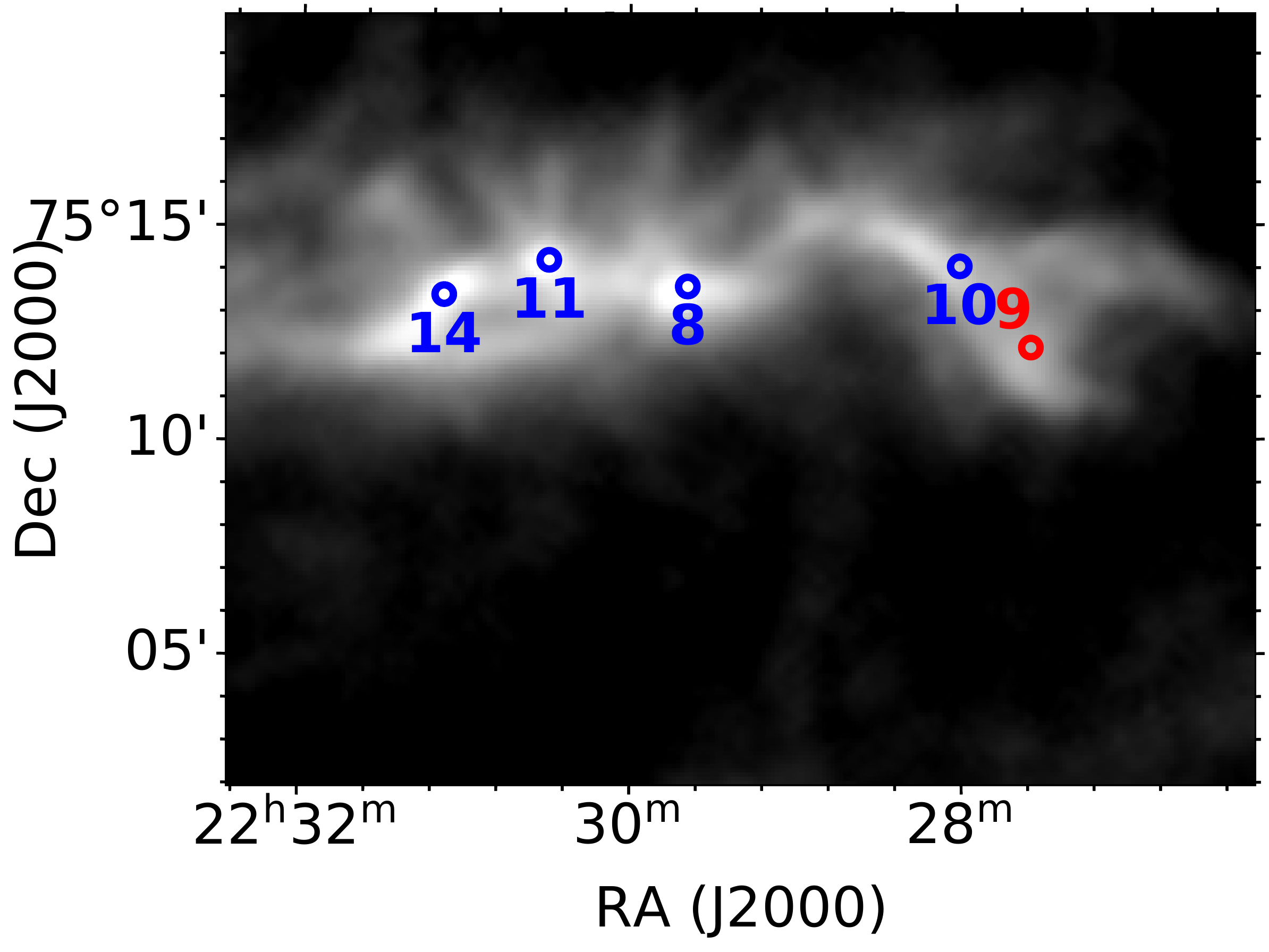}
    \includegraphics[width=68mm]{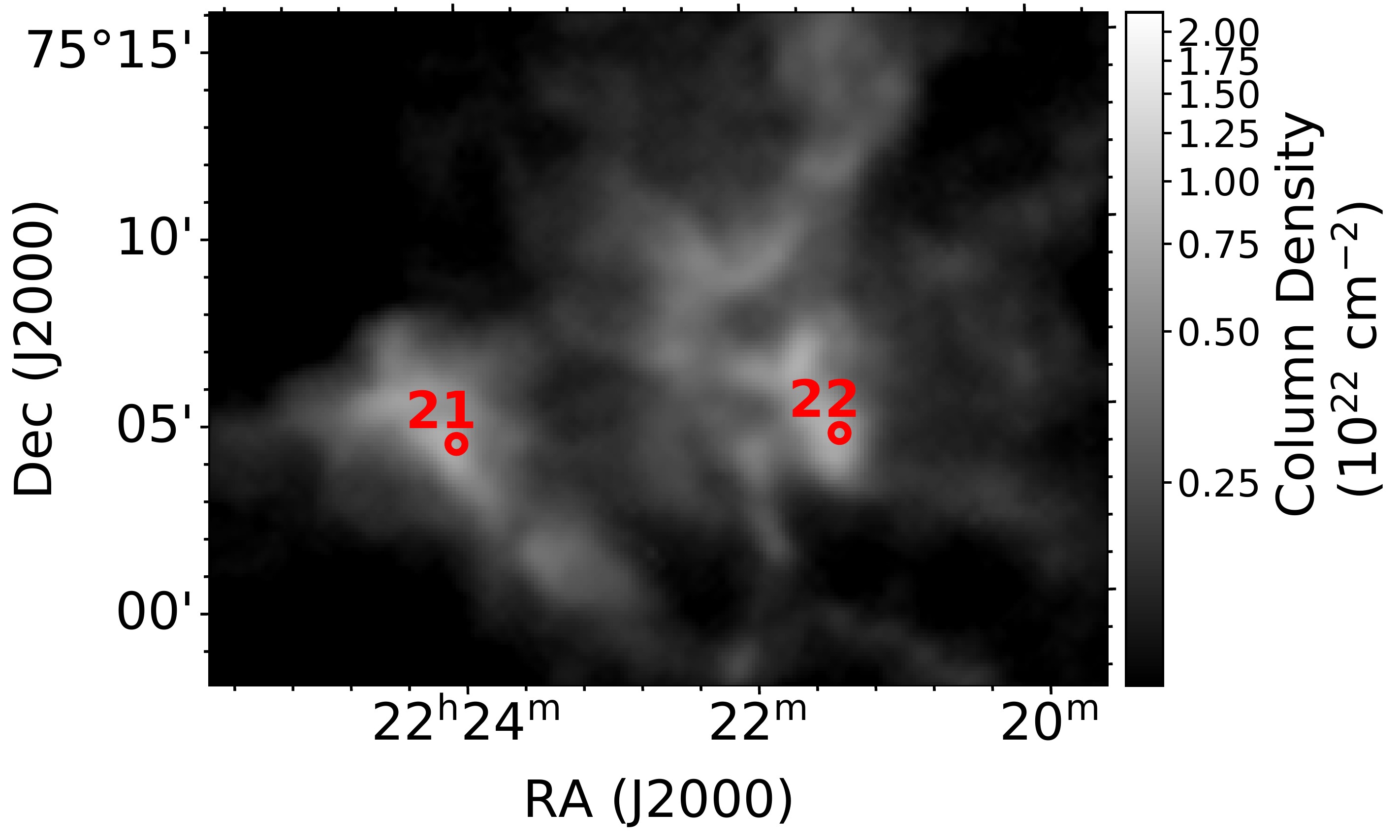}
    \caption{Three regions of L1251 with all twenty-two sources identified (o-NH$_2$D detections shown in blue; non-detections shown in red). Background image is H$_2$ column density from \citealt{2020ApJ...904..172D}.}
    \label{fig:Cepheus}
\end{figure*}

Astrochemical studies typically focus on details of a single object or a small subset of objects located in different clouds. Therefore, there are a lack of surveys that focus on the chemistry of the entire dense core population within a single cloud. 
This work is an observational study of singly-deuterated ammonia, o-NH$_2$D, in the c-type rotation-inversion transition $1_{11}^{+} \rightarrow 1_{01}^{-}$ (optically thin critical density of $7 \times 10^4$ cm$^{-3}$ using the definition in \citealt{2015PASP..127..299S}) toward the entire NH$_3$-identified core population in the nearby Lynds 1251 (L1251) molecular cloud. 
L1251 is an isolated molecular cloud actively forming low-mass stars, located in the near side of the Cepheus Flare \citep{1989ApJ...347..231G, 2008hsf1.book..136K}. 
Maps of $^{13}$CO 1-0 indicate that the cloud has a cometary, filamentary shape spanning over 7 pc in length \citep{1994ApJ...435..279S}.
Twenty-two dense cores, identified by the NH$_3$ survey of \citealt{2017ApJ...850....3K}, were observed using the 12m Arizona Radio Observatory Telescope on Kitt Peak (see Figure \ref{fig:Cepheus}). 
These observations attempted to detect deuterated ammonia (o-NH$_2$D) to subsequently study the chemical evolution of these cores. 
If NH$_2$D is present, the core (whether starless or prestellar) is predicted to be more evolved and more gravitationally bound (\citealt{Friesen_2013}). 
This study encompasses an entire population of dense cores in a single star-forming cloud, which will allow us to compare each core based on the amount of NH$_2$D in cores with presumably similar chemical initial conditions. 

%Figure 1

Section \ref{section:Observations} outlines the observational details and source selections. Section \ref{section:Analysis} is an analysis of derived parameters such as filling fraction, column density, and velocity dispersion. Filling fraction is not assumed to be 1; we discuss its calculation and the ramifications of a non-unity filling fraction. We discuss our results in Section \ref{sec:results} and report the deuterium ratio for each core.
Plots of several physical and evolutionary parameters show variation between cores with NH$_2$D detection and those without.
Chemical evolution simulations were run to understand o-NH$_2$D abundance throughout the lifetime of the core. 
We compare our observed deuteration ratios to chemical models of prestellar deuteration in the literature.

\section{Observations}
\label{section:Observations}

The data was obtained using the Arizona Radio Observatory (ARO) 12 m Telescope on Kitt Peak.
Observations were made with the 3 mm sideband separating dual polarization receiver.
The AROWS spectrometer, with a resolution of 0.1362 km s$^{-1}$, was used for all observations.
The two polarizations (vertical and horizontal) were averaged together. Observations occurred from October 25, 2018 to December 2, 2018.
A total of 22 sources were observed in the o-NH$_2$D $1_{11}^{+} \rightarrow 1_{01}^{-}$ transition at 85.926278 GHz toward the Cepheus L1251 molecular cloud. 
These cores were first mapped in p-NH$_3$ (1,1) and (2,2) by \citealt{2017ApJ...850....3K} and we point towards the peak ammonia (1,1) intensity position.
Figure \ref{fig:Cepheus} shows the surveyed portion of L1251 with all 22 sources. 
We use the shorthand notation "Cxx" to name each source where the "C" means Cepheus and the "xx" is a two digit number that corresponds to the catalog number in \citealt{2017ApJ...850....3K}.
The angular resolution of our 12m observations is 70.16$^{\prime\prime}$.
The telescope pointing corrections are performed every two hours and are typically less than $5^{\prime\prime}$.
We assume the distance to the sources is $351^{+20}_{-19}$ pc from an analysis of stellar photometry data with GAIA parallax distances (including their quoted 5\% systematic error added in quadrature; \citealt{2020A&A...633A..51Z}).
Since \cite{2017ApJ...850....3K} asumed a distance of 300 pc, then all distance dependent quantities reported in their paper have been scaled to the new distance.

Absolute position switching was used during observations. We chose three off positions which were checked to be devoid of o-NH$_2$D emission. Each source was observed for a minimum of $4$ hours in multiples of 5 minute scans; alternating 30 seconds on source and then 30 seconds off source.
The data reductions made use of the CLASS package in GILDAS software \citep{2005sf2a.conf..721P}.
Observations of Jupiter, Mars, or Saturn were used for pointing and focusing every 2 hours during the observations.

\begin{figure}
    \centering
    \includegraphics[width=85mm]{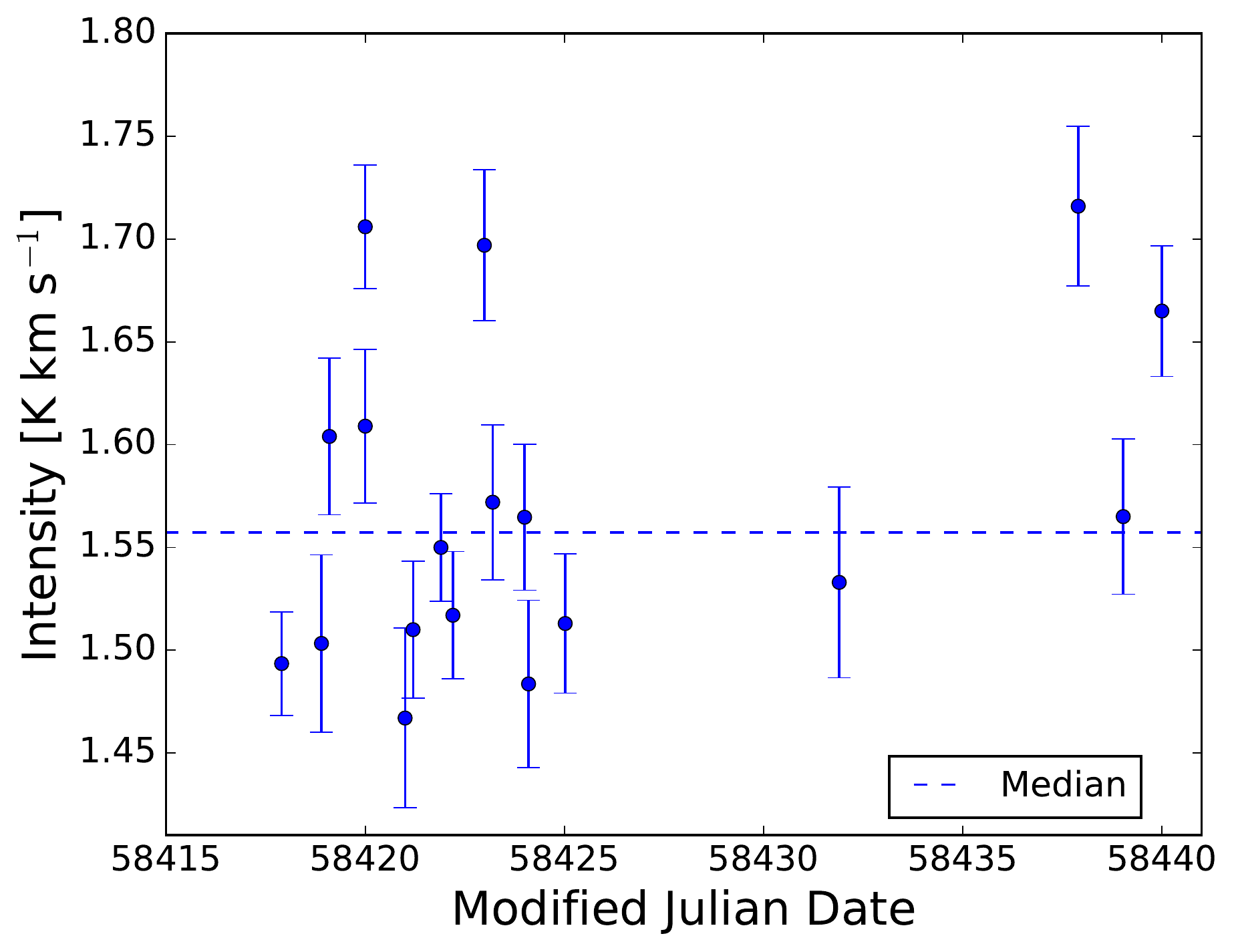}
    \caption{Plot of integrated intensity (T$_A^*$ scale) vs Modified Julian Date for the calibration source L183.}
    \label{fig:Intensity}
    %\vspace{-1cm}
\end{figure}

Spectra obtained at the ARO 12 m telescope are calibrated on the T$_A^*$ scale using the chopper-wheel calibration method (\citealt{1981ApJ...250..341K}).
The consistency of the telescope calibration was checked each shift using observations of the prestellar core L183.
The coordinates of L183 in J2000.0 epoch are 15$^h$ 54$^m$ 08.6$^s$, -02$^{\rm{o}}$ 52$^{\prime}$ 40.0$^{\prime\prime}$.
This source was chosen because of its high integrated intensity in o-NH$_2$D (see Figure \ref{fig:Intensity}) with an average intensity of 1.57 $\pm$ 0.08 K km s$^{-1}$.
The maximum variation observed was $9$\% from the average value.
Observations of Mars were used to calculate the main beam efficiency of the telescope. The polarization-averaged main beam efficiency was calculated to be 0.847 $\pm$ 0.032.  Observations were placed onto the main beam temperature scale (T$_{mb}$) by dividing the observed spectrum by the main beam efficiency, T$_{mb}$ = T$_A^*$/$\eta_{mb}$.
The median baseline rms of the 22 observed cores was $\sigma_{T_{mb}} = 17$ mK.

% Figure 2

%\begin{landscape}
\begin{figure*}
    \centering
    \includegraphics[width=\textwidth]{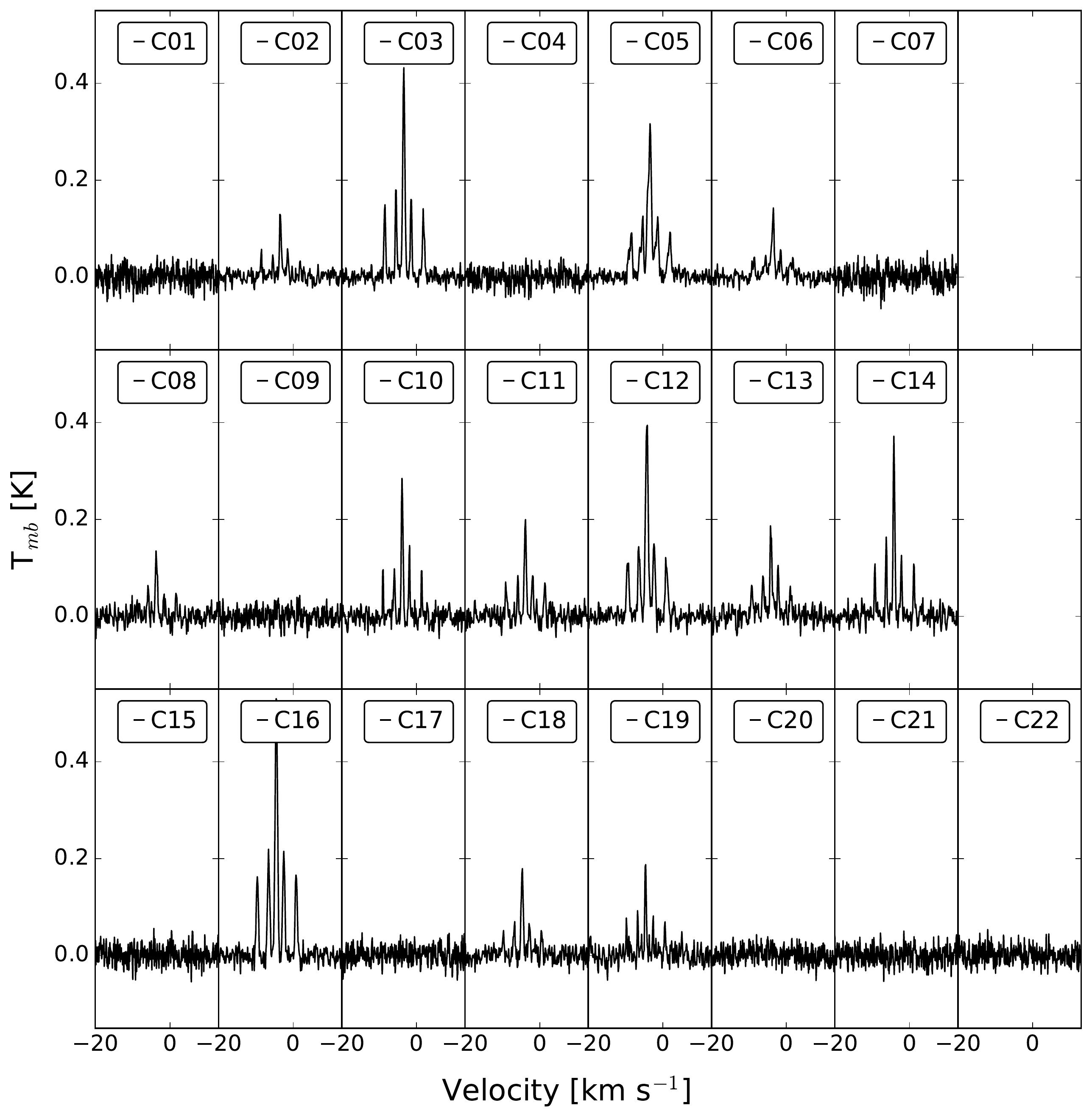}
    \caption{o-NH$_2$D $1_{11}^{+} \rightarrow 1_{01}^{-}$ spectra of all twenty-two sources in order. The y-axis is the $T_{mb}$ scale and the x-axis is $v_{LSR}$.}
    \label{fig:22spec}
\end{figure*}
%\end{landscape} 

\subsection{Starless vs. Protostellar Classification}

The vast majority (20 out of 22) of NH$_3$-identified cores in L1251 are characterized as starless or prestellar.
We checked for far-infrared point sources by comparing the \textit{Herschel} $70$ $\mu$m image with the NH$_3$ core positions \citep{2020ApJ...904..172D}.
Bright $70$ $\mu$m emission can be expected from the warm dust surrounding Class 0 and I protostars \citep{2014prpl.conf..195D}.
Only two cores, C05 and C11, have clearly associated $70$ $\mu$m point sources that were within the  GBT beam ($32^{\prime\prime}$) of the NH$_3$ (1,1) intensity peak.
C05 is most closely associated with the $0.8$ L$_{\odot}$ Class 0/I protostar L1251B-IRS1, while two more Class I protostars, L1251B-IRS2 and L1251B-IRS4 are projected near the core edge \citep{2006ApJ...648..491L}.
C11 contains an $0.8$ L$_{\odot}$ Class 0 protostar, L1251A-IRS3, with a long, collimated infrared jet \citep{2010ApJ...709L..74L}.
We classify C05 and C11 as protostellar.

There were two cases with close associations where a $70$ $\mu$m source was near the edge of the NH$_3$ core making their classification less certain.
C14 is $60^{\prime\prime}$ from a $70$ $\mu$m source to its northwest that is associated with the $0.6$ L$_{\odot}$ Class 0 protostar L1251A-IRS4 \citep{2010ApJ...709L..74L}.
It is not clear that this protostar is directly associated with the NH$_3$ core.
A $70$ $\mu$m source is also found in between C06 and C12 which is a Class II protostar \citep{2006ApJ...648..491L}.
We classify C06 and C12 as starless.

We list whether the core is associated with a protostar in Table \ref{tab:SourceProp}.
All five cores associated with or nearby $70$ $\mu$m point sources are detected in o-NH$_2$D $1_{11}^{+} \rightarrow 1_{01}^{-}$, which may imply a relation between the presence of a $70$ $\mu$m source and more evolved chemistry. 
We do not analyze the two high confidence protostellar sources as a separate subset in the remainder of the paper due to the small sample size.

\begin{table*}
\centering
\caption{NH$_3$-identified core physical properties reported in and derived from \citealt{2017ApJ...850....3K}, scaled to the new distance of 351 pc.  Coordinates are epoch J2000.0.}
\label{tab:SourceProp}
\begin{tabular}{lllcllll}
\hline
Core & RA                   & Dec                     & 70 $\mu$m & Mass          & Radius         & $\bar{n}$        & T$_K$          \\
     & $\alpha$             & $\delta$                &           & M$_{\odot}$   & 0.01 pc   & 10$^4$ cm$^{-3}$ & K              \\
     \hline
1    & 22$^h$36$^m$41.6$^s$ & 75$^{\circ}$08' 22.1'' & N         & 2.7 $\pm$ 0.5 & 6.5 $\pm$ 0.8  & 3.0 $\pm$ 0.6   & 11.0 $\pm$ 1.7   \\
2    & 22$^h$37$^m$36.7$^s$ & 75$^{\circ}$11' 44.0'' & N         & 3.8 $\pm$ 0.8 & 5.3 $\pm$ 0.7  & 7.0 $\pm$ 1.6   & 10.5 $\pm$ 0.6 \\
3    & 22$^h$39$^m$02.9$^s$ & 75$^{\circ}$11' 46.6'' & N         & 2.7 $\pm$ 0.5 & 2.2 $\pm$ 1.1  & 34.8 $\pm$ 18.3 & 11.3 $\pm$ 0.5 \\
4    & 22$^h$37$^m$41.4$^s$ & 75$^{\circ}$09' 46.2''  & N         & 4.2 $\pm$ 0.8 & 7.8 $\pm$ 1.1  & 2.7 $\pm$ 0.5   & 10.3 $\pm$ 1.6 \\
5    & 22$^h$38$^m$47.5$^s$ & 75$^{\circ}$11' 14.1'' & Y         & 3.1 $\pm$ 0.7 & 1.6 $\pm$ 1.4  & -         & 13.1 $\pm$ 0.7 \\
6    & 22$^h$39$^m$17.3$^s$ & 75$^{\circ}$10' 34.8'' & N         & 2.1 $\pm$ 0.4 & 2.2 $\pm$ 1.1  & 25.6 $\pm$ 13.1 & 10.7 $\pm$ 0.5 \\
7    & 22$^h$38$^m$00.9$^s$ & 75$^{\circ}$11' 39.7'' & N         & 1.5 $\pm$ 0.3 & 3.1 $\pm$ 0.8  & 9.8 $\pm$ 2.7   & 9.4 $\pm$ 0.7  \\
8    & 22$^h$29$^m$40.5$^s$ & 75$^{\circ}$13' 32.0'' & N         & 7.5 $\pm$ 1.5 & 6.8 $\pm$ 1.0    & 7.1 $\pm$ 2.5   & 9.5 $\pm$ 0.6  \\
9    & 22$^h$27$^m$35.2$^s$ & 75$^{\circ}$12' 05.4''  & N         & 2.3 $\pm$ 0.4 & 5.3 $\pm$ 0.9  & 4.1 $\pm$ 0.9   & 8.9 $\pm$ 1.3  \\
10   & 22$^h$28$^m$01.0$^s$ & 75$^{\circ}$13' 59.6'' & N         & 10.0 $\pm$ 2.1  & 11.5 $\pm$ 1.1 & 2.2 $\pm$ 0.7   & 9.1 $\pm$ 0.9  \\
11   & 22$^h$30$^m$31.2$^s$ & 75$^{\circ}$14' 08.6'' & Y         & 7.1 $\pm$ 1.4 & 7.1 $\pm$ 0.9  & 6.0 $\pm$ 2.1   & 9.8 $\pm$ 0.5  \\
12   & 22$^h$39$^m$30.0$^s$ & 75$^{\circ}$11' 00.7'' & N         & 3.8 $\pm$ 0.8 & 2.9 $\pm$ 1.0    & 28.7 $\pm$ 10.4 & 10.0 $\pm$ 0.6   \\
13   & 22$^h$38$^m$29.4$^s$ & 75$^{\circ}$11' 25.2'' & N         & 2.5 $\pm$ 0.5 & 3.2 $\pm$ 0.9  & 15.0 $\pm$ 4.8  & 11.6 $\pm$ 0.4 \\
14   & 22$^h$31$^m$09.5$^s$ & 75$^{\circ}$13' 19.7'' & N?        & 12.0 $\pm$ 2.5  & 9 $\pm$ 1.0      & 5.3 $\pm$ 1.9   & 9.9 $\pm$ 0.4  \\
15   & 22$^h$35$^m$24.6$^s$ & 75$^{\circ}$18' 52.8'' & N         & 3.7 $\pm$ 0.7 & 5.7 $\pm$ 0.8  & 5.5 $\pm$ 1.6   & 12.5 $\pm$ 1.6 \\
16   & 22$^h$39$^m$38.4$^s$ & 75$^{\circ}$11' 48.3'' & N         & 4.5 $\pm$ 1.0   & 3.1 $\pm$ 0.9  & 31.1 $\pm$ 11   & 9.5 $\pm$ 0.3  \\
17   & 22$^h$35$^m$03.5$^s$ & 75$^{\circ}$13' 01.9'' & N         & 1.2 $\pm$ 0.3 & 4.4 $\pm$ 0.8  & 3.8 $\pm$ 0.9   & 10.3 $\pm$ 1.7 \\
18   & 22$^h$39$^m$20.2$^s$ & 75$^{\circ}$13' 29.1'' & N         & 6.6 $\pm$ 1.4 & 6.2 $\pm$ 1.0    & 7.8 $\pm$ 2.3   & 9.4 $\pm$ 1.3  \\
19   & 22$^h$35$^m$52.9$^s$ & 75$^{\circ}$18' 46.8'' & N         & 10.0 $\pm$ 2.1  & 9.1 $\pm$ 1.0    & 4.3 $\pm$ 1.1   & 10.5 $\pm$ 0.8 \\
20   & 22$^h$35$^m$33.5$^s$ & 75$^{\circ}$16' 57.5'' & N         & 3.1 $\pm$ 0.7 & 3.8 $\pm$ 1.2  & 13.1 $\pm$ 4.3  & 12.9 $\pm$ 2.7 \\
21   & 22$^h$24$^m$04.0$^s$ & 75$^{\circ}$04' 19.5'' & N         & 1.8 $\pm$ 0.4 & 4.9 $\pm$ 1.2  & 3.9 $\pm$ 1.1   & 10.0 $\pm$ 1.9   \\
22   & 22$^h$21$^m$25.0$^s$ & 75$^{\circ}$04' 20.3'' & N         & 1.9 $\pm$ 0.4 & 5.0 $\pm$ 0.9    & 3.9 $\pm$ 0.8    & 10.5 $\pm$ 1.7
\end{tabular}
\end{table*}

\section{Spectral Analysis}
\label{section:Analysis}

o-NH$_2$D emission was detected in 13 (59\%) of the NH$_3$-detected cores identified by \cite{2017ApJ...850....3K}. 
For the analysis of the hyperfine structure of o-NH$_2$D emission, we use the splitting due to both the N and D atoms ($\vec{F}_1 = \vec{J} + \vec{I}_N$ and $\vec{F} = \vec{F}_1 + \vec{I}_D$; \citealt{2016A&A...586L...4D}) to designate two hyperfine quantum numbers for each hyperfine energy level ($F_1$, $F$).
See Appendix \ref{section:AppHyp} for more details.
All spectral line data were analyzed using the GILDAS CLASS software hyperfine fitting routines to determine the total optical depth ($\tau$), line centroid ($V_{\rm{LSR}}$), and FWHM line width ($\Delta v$) and the results are presented in Table \ref{tab:HyperFits}.
The CLASS fitting routine assumes hyperfine statistical equilibrium among the hyperfine transitions (see \citealt{2010ApJ...716.1315K}).
Unfortunately, for 9 of the 13 detections, the signal-to-noise (SNR) of the spectrum was not sufficient ($3\sigma_{\tau} > \tau$) to accurately constrain the optical depth.
In this section, we analyze the hyperfine fit parameters, correct for the source-beam filling fraction, and calculate the column densities of o-NH$_2$D.

\begin{table*}
\centering
\caption{Properties of $1_{11}^{+} \rightarrow 1_{01}^{-}$ o-NH$_2$D spectra.}
\label{tab:HyperFits}
\begin{tabular}{llllllllll}
\hline
Core & $\sigma_{T_A^*}$ & $I(T^*_A)$     & $\sigma_{I(T^*_A)}$ & V$_{lsr}$   & $\sigma_{Vlsr}$ & $\Delta v$  & $\sigma_{\Delta v}$ \\
     & mK              & mK km s$^{-1}$ &   mK km s$^{-1}$                  & km s$^{-1}$ &    km s$^{-1}$             & km s$^{-1}$ &     km s$^{-1}$                \\
\hline\hline
1    & 16          &  -              & 14                & -           & -               & -           & -                   \\
2    & 14          & 131            & 9                 & -3.499      & 0.012           & 0.34       & 0.04               \\
3    & 15          & 565            & 10                  & -3.460       & 0.005           & 0.45       & 0.01               \\
4    & 13          &   -             & 11                & -           & -               & -           & -                   \\
5    & 17          & 646            & 9                 & -      & -           & -        & -               \\
6    & 15          & 231            & 9                 & -3.637      & 0.021           & 0.82       & 0.06               \\
7    & 15          &   -             & 14                & -           & -               & -           & -                   \\
8    & 14          & 141            & 12                & -3.721      & 0.015           & 0.41        & 0.05               \\
9    & 16          &   -             & 11                & -           & -               & -           & -                   \\
10   & 15          & 264            & 13                & -3.914      & 0.006           & 0.26       & 0.02               \\
11   & 14          & 222            & 13                & -3.916      & 0.009           & 0.42       & 0.02               \\
12   & 12          & 653            & 11                & -4.393      & 0.005           & 0.71       & 0.01               \\
13   & 15          & 234            & 14                & -4.172      & 0.010            & 0.40       & 0.03                \\
14   & 15          & 351            & 13                & -4.201      & 0.004           & 0.32       & 0.01               \\
15   & 14          &   -             & 13                & -           & -               & -           & -                   \\
16   & 14          & 814            & 13                & -4.578      & 0.004           & 0.58       & 0.01               \\
17   & 16          &    -            & 13                & -           & -               & -           & -                   \\
18   & 15          & 207            & 13                  & -4.778      & 0.011           & 0.47       & 0.03               \\
19   & 17          & 200            & 15                & -4.729      & 0.008           & 0.23       & 0.02               \\
20   & 13          &     -           & 11                & -           & -               & -           & -                  \\
21   & 15          &     -           & 13                & -           & -               & -           & -                   \\
22   & 14          &     -           & 12                & -           & -               & -           & -                                    
\end{tabular}
\end{table*}

\subsection{Velocity Centroid and Velocity Dispersion}

\begin{figure}
    \centering
    \includegraphics[width=85mm]{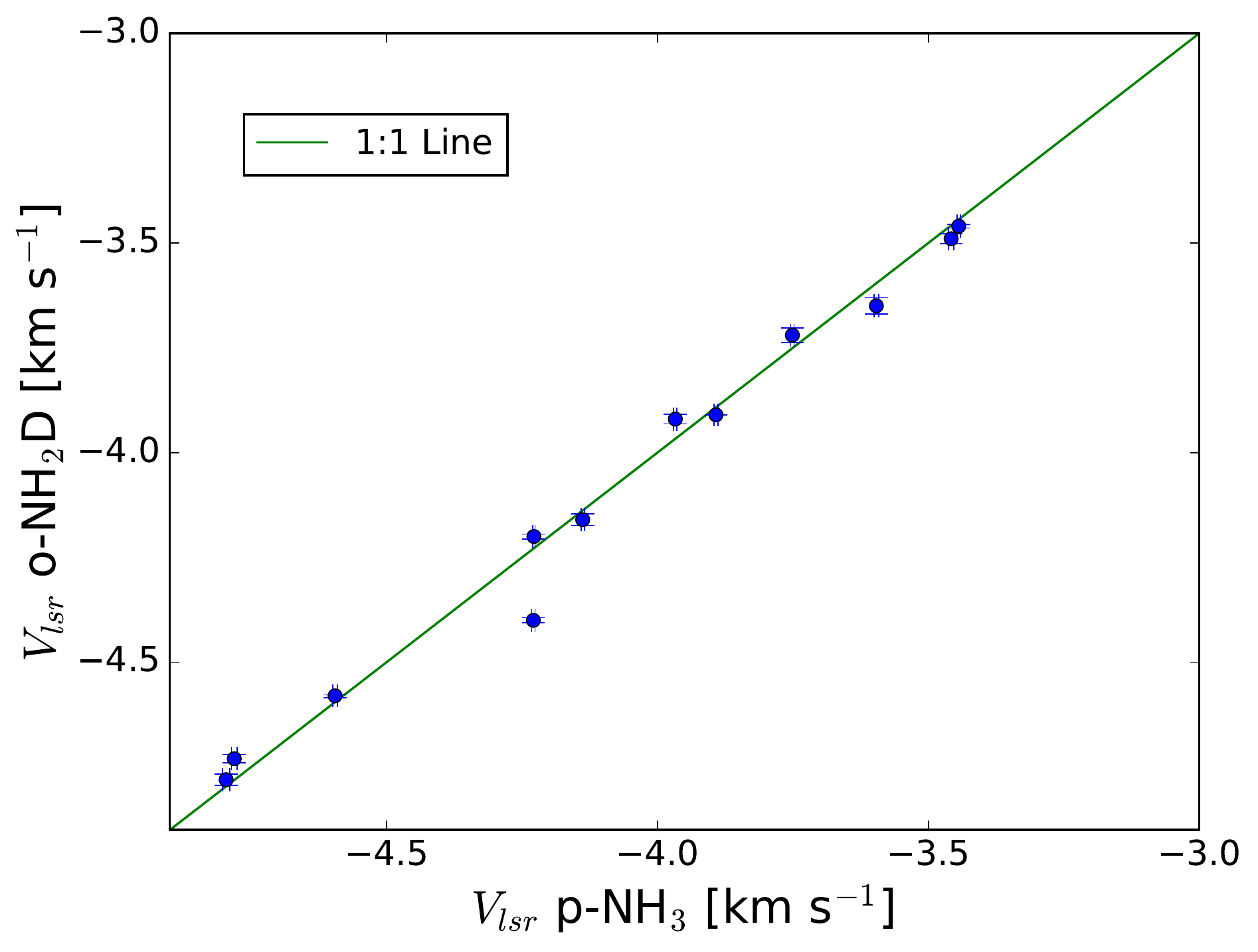}
    \caption{Comparison between o-NH$_2$D $1_{11}^{+} \rightarrow 1_{01}^{-}$ and p-NH$_3$ (1,1) radial velocities. The green line corresponds to equal V$_{LSR}$. The median difference in V$_{LSR}$ is 0.029 km s$^{-1}$. We exclude C05 from this plot and these calculations. The source plotted with the largest offset is C12.}
    \label{fig:vlsr}
\end{figure}

The velocity centroid for both o-NH$_2$D $1_{11}^{+} \rightarrow 1_{01}^{-}$ and p-NH$_3$ (1,1) are well determined from fitting the hyperfine pattern, even for low SNR spectra, because of the multiple hyperfine transitions.
All sources, with the exception of C05, are well fit by a single velocity component (Figure \ref{fig:22spec}).
There are two velocity components evident in the spectrum of C05. We shall exclude this source from the following discussion.
The o-NH$_2$D $1_{11}^{+} \rightarrow 1_{01}^{-}$ V$_{\rm{LSR}}$ and the p-NH$_3$ (1,1) V$_{\rm{LSR}}$ are compared in Figure \ref{fig:vlsr}. 
There is excellent agreement between the o-NH$_2$D  and the p-NH$_3$ (1,1) velocity centroids with an average ratio of 1.0 and a median difference between V$_{LSR}$ of 0.029 km s$^{-1}$. The source with the biggest V$_{LSR}$ difference is C05, which has multiple velocity components and is disregarded in later calculations. The source with the second biggest difference is C12, which may be potentially multi-peaked (unresolved) and has the second largest linewidth.
Even though the ARO 12 m beamsize is $2.1$ times larger in angular diameter than the GBT beamsize, this result indicates that the o-NH$_2$D emission in the larger beam is not strongly affected by gradients in the core kinematics compared to NH$_3$ observed in the smaller beam (cf. \citealt{2019ApJ...886..119C}).

The FWHM line width was determined from the hyperfine fits assuming every hyperfine transition has the same FWHM line width.
The mean FWHM line width for o-NH$_2$D $1_{11}^{+} \rightarrow 1_{01}^{-}$ is $\Delta v = 0.50 \pm 0.24$ km s$^{-1}$ and the median value is $0.42$ km s$^{-1}$.
We note that the cores with the largest o-NH$_2$D line widths (excluding the multi-component source C05) are C06 ($0.82 \pm 0.06$ km s$^{-1}$), and C12 ($0.71 \pm 0.01$ km $s^{-1}$) hinting that the dense gas in these cores is possibly affected by nearby protostellar activity.

The thermal and non-thermal components of velocity dispersion were calculated from 
\begin{equation}
    \sigma^2 = \sigma_{nt}^2 + \sigma_{th}^2 = \left[ \frac{\Delta v^2}{8ln(2)} - \frac{k T_k}{\mu_{mol} m_H} \right] + \frac{k T_k}{\mu_p m_H}
\end{equation}
where $\Delta v$ is the observed FWHM line width, T$_k$ is the gas kinetic temperature (from \citealt{2017ApJ...850....3K} NH$_3$ (1,1) and (2,2) observations), $k$ is the Boltzmann constant, $\mu_{mol} = 17$ amu  for the molecular mass of NH$_3$ and $18$ amu for the molecular mass of NH$_2$D, $m_H$ is the mass of hydrogen, and $\mu_p = 2.34$ is the mean mass of a gas particle \citep{1992ApJ...384..523F}.

The average nonthermal velocity dispersion of o-NH$_2$D is $\sigma_{nt} = 0.18 \pm 0.08$ km s$^{-1}$, while the average thermal velocity dispersion (equal to the sound speed) is $\sigma_{th} = 0.19 \pm 0.01$ km s$^{-1}$. 
The average ratio of non-thermal to thermal dispersion is $\frac{\sigma_{th}}{\sigma_{nt}} = 0.93 \pm 0.42$ with a median ratio of $0.88$.
This ratio implies an almost equal amount of thermal to nonthermal support in each core detected on average.

The ratio between the o-NH$_2$D non-thermal velocity dispersion and the p-NH$_3$ non-thermal velocity dispersion (from an NH$_3$-intensity weighted average over the source; \citealt{2017ApJ...850....3K}) is $1.16 \pm 0.29$ for the subset of 12 sources with single velocity components that are detected in both lines. 
Again, this ratio is consistent with the o-NH$_2$D emission not being strongly affected by gradients in the core kinematics compared to p-NH$_3$ on average even though the 12m beam is larger than the GBT beam.  
There are, however, individual cases where the ratio is as small as $0.63$ (C19) and as large as $1.64$ (C06).
There is no statistical difference in the observed FWHM line width of the p-NH$_3$ (1,1) transition between o-NH$_2$D-detected cores and o-NH$_2$D-non-detected cores ($\Delta v = 0.40 \pm 0.11$ km s$^{-1}$ vs. $\Delta v = 0.35 \pm 0.12$ km s$^{-1}$, calculated from the observed velocity dispersion in \citealt{2017ApJ...850....3K}).

\subsection{Source Size \& Filling Fraction}
\label{section:fillingfraction}

The filling fraction, $f$, measures the coupling between the source integrated intensity emission ($I(\theta,\phi)$) and the telescope beam ($P_n(\theta)$).
If we assume that the source integrated intensity distribution on the sky can be approximated by a Gaussian with FWHM $\theta_{src}$, then the filling fraction is given by
\begin{equation}
     f =     \frac{\int I(\theta,\phi) P_n(\theta) d\Omega}{I(\theta = 0)\int P_n(\theta) d\Omega} = 
     \frac{\theta_{src}^2}{\theta_{src}^2 + \theta_{beam}^2}  \;\; ,
\end{equation}
where $\theta_{beam}$ is the ARO 12 m beam size (70.16$^{\prime\prime}$). 
For example, if the source FWHM matches the telescope beam FWHM, then $f = 0.5$ and the observed intensity is reduced by a factor of 0.5 due to the source-beam coupling.
We calculated the filling fraction for each source and applied the correction to the observed main beam line temperature, $T_R = T_{mb}/f$.

We determine the NH$_3$ (1,1) half-peak intensity of each source to calculate the source size used for the filling fraction calculations. 
This is done by finding the solid angle, $\Omega_{\rm{fwhm}}$ (square arcseconds), of the contour at half-peak intensity.  
The uncertainty on the solid angle is calculated from the contours corresponding to $\pm 1\sigma$ error in the p-NH$_3$ (1,1) integrated intensity of 0.154 K km/s. 
The angular source size is determined by converting $\Omega_{\rm{fwhm}}$ into the angular diameter of a circle with the same solid angle and then by de-convolving the GBT telescope beam size ($32^{\prime\prime}$),
$\theta_{src}^2 = \frac{4 \Omega_{\rm{fwhm}}}{\pi} - {32}^2$.
The filling fraction in the ARO 12m telescope beam is reported in Table \ref{tab:ColDen}.
The average filling fraction found using Equation 2 across all 22 sources was found to be f$ = 0.38 \pm 0.12$ with a median value of $0.38$. 
Because of the close proximity of some of the sources, the half-peak intensity contour may encompass multiple sources. 
In these cases, we assign the median filling fraction of 0.38 (see \S3.3 for a different procedure for the optically thick sources). 
We also calculate the filling fraction for the GBT beam using the same method and find an average filling fraction f$_{GBT}$ = 0.82.

For any other calculations that use radius, we use the p-NH$_3$ (1,1) dendrogram leaves to estimate the source size.
These were calculated using p-NH$_3$ (1,1) integrated intensity maps and summing over the pixels identified within the source from the dendrogram analysis presented in \citealt{2017ApJ...850....3K}.
The source size was then calculated from the radius of a circle that has the same area as the dendrogram leaf and de-convolving the GBT beam size.
This source size is used for all calculations in the paper that utilize the radius of the core.
Errors in the dendrogram-derived source size were estimated by summing the perimeter pixels and assuming an error of $\pm$ 0.5 pixels. 
This dendrogram-derived radius and error are reported in Table \ref{tab:SourceProp}.

\subsection{Column Densities} \label{section:columndensity}

o-NH$_2$D column densities were calculated using the CTEX method (CTEX = constant excitation temperature approximation) where the same excitation temperature, $T_{ex}$, is assumed for all energy levels in a Boltzmann distribution (see Appendix A of \citealt{2002ApJ...565..344C} and \citealt{2015PASP..127..266M}).
The CTEX column density was calculated from  the strongest hyperfine component ($J^{\pm}_{K_a,K_c},F_1, F = $ $1^+_{1,1}$, 2, 3 $-$ $1^-_{0,1}$, 2, 3) by assuming a Gaussian line profile with FWHM line width, $\Delta v$, from the equation
\begin{equation}
N(\rm{o-NH}_2\rm{D}) = 
\frac{R_i}{h g_{u,i} A_i} \frac{Q_{hyp}(T_{ex}) \, u_{\nu_i}(T_{ex})}{\exp(-E_{u,i}/kT_{ex})}\, \sqrt{\frac{\pi}{4 \ln(2)}} \, \tau \Delta v 
\end{equation}
where $R_i = 14/81$ is hyperfine relative intensity (see Appendix \ref{section:AppHyp}), 
$g_{u,i} = 2F + 1 = 7$ is the statistical weight of the upper hyperfine energy level, 
$A_i = 5.23 \times 10^{-6}$ s$^{-1}$ is the hyperfine transition spontaneous emission coefficient \citep{2016A&A...586L...4D}, 
$Q_{\rm{hyp}}$ is the partition function summed over all of the o-NH$_2$D $J_{K_a,K_c}^{\pm} F_1 F$ energy levels, 
$u_{\nu}(T) = (8\pi h \nu^3/c^3)/[\exp(h\nu / kT) - 1]$ is the Planck energy density (erg cm$^{-3}$ Hz$^{-1}$),
$E_{u,i}/k = 20.0894$ K is the upper energy level of the hyperfine transition above the ground rotation-inversion level of o-NH$_2$D ($0^-_{0,0}$),
and $\tau$ is the total line optical depth summed over all hyperfine transitions. 
The excitation temperature is determined from the velocity integrated radiative transfer equation for position-switched observations (see \S3 of \citealt{2015PASP..127..266M}),
\begin{equation}\label{eq:excitationtemp}
T_{ex} =  \frac{h \nu}{k} \left[ \ln \left( 1 + \frac{\frac{h \nu}{k}}{\frac{\frac{h \nu}{{k}}}{\exp{(h\nu /k T_{cmb})} - 1} + \frac{\int T_R dv}{W_v}} \right) \right] ^{-1} 
\end{equation}
where $W_v = \int 1 - \exp(-\tau(v)) \, dv$ is the equivalent width (km s$^{-1}$) integrated over the entire theoretical hyperfine pattern for a given optical depth and FWHM line width, assuming a Gaussian line profile for each hyperfine line.
For optically thin emission lines, there is a degeneracy between the optical depth and the excitation temperature meaning that $T_{ex}$ must be assumed to determine a column density.
Optically thick emission lines do not have this degeneracy but their analysis requires an accurate determination of the optical depth for an accurate determination of the column density.
One difficulty with deriving optical depth from hyperfine fitting is that an accurate determination of the total optical depth requires very high signal-to-noise ratio spectra.
Only 3 cores (C03, C12, and C16) were found to be optically thick with reliable fits ($\tau > 1$ and $3\sigma_{\tau} < \tau$) from the CLASS hyperfine fitting.
We shall use these three sources to constrain a plausible range in $T_{ex}$ to be applied to all sources.
An additional complication is that all three reliably optically thick sources do not have well constrained filling fractions, therefore we must find an alternative method for simultaneously constraining this quantity.

In order to constrain the filling fraction of the optically thick sources and ultimately their excitation temperature, we compared the results between the CTEX method and an alternative method of determining the column density using RADEX, a constant density, constant temperature non-LTE statistical equilibrium radiative transfer code \citep{2007A&A...468..627V}.
We chose the escape probability formalism for a homogeneous, static spherical medium.
The standard RADEX inputs are the gas kinetic temperature ($T_k$), average volume density ($\bar{n}$), molecular column density ($N(\rm{o-NH}_2\rm{D})$), molecular FWHM line width ($\Delta v$), and a molecular collision rate file determined from quantum-collision calculations (see  \citealt{2014MNRAS.444.2544D} for the collision rate calculation for o-NH$_2$D with p/o-H$_2$).  
RADEX then calculates the excitation temperature of a transition ($T_{ex}$), the line-center optical depth ($\tau$), and the integrated intensity ($I = \int T_R dv$).
Collision rate coefficients do not currently exist in the literature for the separate hyperfine energy levels of o-NH$_2$D, therefore the integrated intensity is calculated over the entire hyperfine pattern.
The RADEX model can then be compared to the observed integrated intensity ($T_R$ scale) to constrain the column density.
We use the integrated intensity, $I(T_R)$, and the FWHM line width from the hyperfine fitting of the spectrum. 
The gas-kinetic temperature is derived from the NH$_3$ observations in \citealt{2017ApJ...850....3K}.
The average density was determined from the average H$_2$ column density ($N(H_2)$) over the core, derived from \textit{Herschel Space Observatory} observations of optically thin dust continuum emission \citep{2020ApJ...904..172D}, and from the beam-deconvolved dendrogram-derived radius of the core ($R$; see \S3.2) assuming a spherical geometry
\begin{equation}
  \label{eq:avgdensity}
  \bar{n} = \frac{3 N(H_2)}{4 R} \;.
\end{equation}

\begin{figure}
    \centering
    \includegraphics[width=90mm]{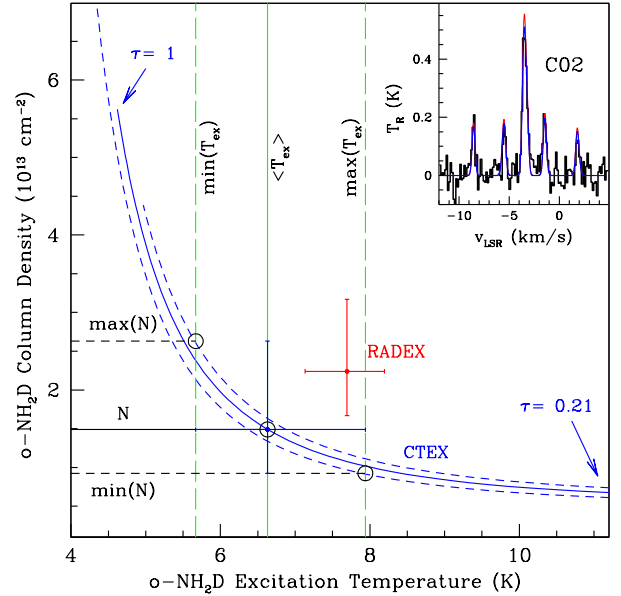}
    \vspace{-2.5cm}
    \caption{An example of the column density calculation for the source C02.  The y-axis is the o-NH$_2$D column density and the x-axis is the o-NH$_2$D excitation temperature.  The blue solid curve shows the column density and excitation temperature combinations that match the observed integrated intensity (on the $T_R$ scale) in the CTEX approximation.  The dashed blue lines show the upper and lower $1 \sigma$ limits in the CTEX approximation.  The optical depth varies along each curve from $\tau = 1$ on the left to $\tau = 0.21$ on the right.  The dashed green lines indicate the $T_{ex}$ range derived from optically thick sources used to constrain the CTEX column density of C02 at the positions they intersect the blue curves (shown as black circles).  The red data point shows the result from the best-fit RADEX models.  The inset shows the observed spectrum with the CTEX model (solid blue curve at $T_{ex} = 6.63$ K) and RADEX model (red) overlaid.}

    \label{fig:RadexCtex}
\end{figure}

We modified a \textit{Python} script that comes with the RADEX source code distribution to vary the column densities until the observed integrated intensity was matched while accounting for all uncertainties on the model input quantities ($T_k$, $\bar{n}$, $\Delta v$) and the observed integrated intensity, $I(T_R)$.
By matching the integrated intensity for the subset of non-optically thick sources, RADEX calculates $T_{ex}$ that ranges between $5.6$ K to $9.0$ K.
This gives us a starting range for plausible excitation temperatures that we can use in the CTEX calculation.
For the three reliably optically thick sources (C03, C12, and C16), the filling fraction was varied until the CTEX and RADEX column densities matched.
The average $T_{ex} = 6.6$ K from the CTEX method at these filling fractions.
The minimum and maximum $T_{ex}$ of $5.7$ K and $8.0$ K was determined by matching the column density $1\sigma$ limits (accounting for all statistical uncertainties in the both the CTEX and RADEX methods).
This CTEX-derived $T_{ex}$ range is comparable to the $T_{ex}$ range calculated by RADEX.
With these numbers determined, an example of the procedure to determine CTEX column densities for the non-optically thick sources is shown in Figure \ref{fig:RadexCtex}.
The blue curves (with dashed curves indicating $1 \sigma$ errors) plot the allowed values of column density and optical depth that match the observed integrated intensity of the source for each $T_{ex}$.
The excitation temperature and limits are then applied (dashed green lines) and where they intersect the CTEX column density curves give the column density and lower and upper limits.
The inset spectrum also shows that the blue CTEX-derived spectrum matches the observed spectrum of the source well.

%Fig CTEX Example

For comparison, in Figure \ref{fig:RadexCtex}, we also plot the RADEX-derived column density and excitation temperature.
While the column density and excitation temperature errorbars overlap significantly between the two different methods,
RADEX is rarely able to simultaneously match the observed integrated intensity, linewidth, optical depth, and excitation temperature (i.e. a RADEX simulated spectrum is often a poor fit to the hyperfine pattern if hyperfine statistical equilibrium is assumed at the RADEX output excitation temperature).
It is for this reason that we do not use the RADEX method to calculate the column densities of all sources and instead use the CTEX method.
One possibility for the RADEX method failure is that the density and temperature of the gas probed by the o-NH$_2$D emission is not equal to the average density and temperature we input into RADEX calculation from the NH$_3$ (1,1) dendrogram-derived radii and NH$_3$ derived gas kinetic temperature.
Using radii and densities derived from the NH$_3$ (1,1) half-peak intensity do not solve this discrepancy.
If we were to make the density and temperature inputs free parameters in RADEX, then new degeneracies (for instance between density and column density) arise which make it difficult to constrain the column density.
Breaking these degenercies would require observations of multiple transtions of o-NH$_2$D.

We report CTEX-derived column densities in Table \ref{tab:ColDen}.
Column density upper limits for non-detections were calculated with the CTEX method for an integrated intensity of $3\sigma_{I(T_R)}$ and assuming the median FWHM linewidth of detected sources, $0.42$ km s$^{-1}$, and the lower limit on $T_{ex} = 5.7$ K.

\begin{landscape}
\begin{table}
\centering
\caption{o-NH$_2$D filling fraction ($f$), column densities, and deuterium fraction.
The excitation temperature, optical depth and o-NH$_2$D column densities are from CTEX model fits. 
Table 3 quantities: f=filling fraction, T$_{ex}$=excitation temperature, $\tau$= total optical depth, N$_{o-NH_2D}$ = ortho-deuterated ammonia column density, N$_{p-NH_3}$ = para-ammonia column density, R$_D$=$\frac{[o-NH_2D]}{[p-NH_3]}$ (deuteration ratio). The$\sigma_f$ and uncertainty on N(p-NH$_3$) are $1\sigma$ error for the filling fraction and ammonia column density. The $\sigma$s for all other quantities are the values calculated with min$\{T_{ex}\}$ and max$\{T_{ex}\}$ values and therefore represent the systematic uncerainty and not a $1\sigma$ statistical uncertainty. Cores with filling fraction values denoted with * use the median filling fraction of 0.38.
}
\label{tab:ColDen}
\begin{tabular}{lllllllllllllllll}
\hline

Core & f & $+\sigma_f$ & $-\sigma_f$ & T$_{ex}$ & $+\sigma_{T_{ex}}$ & $-\sigma_{T_{ex}}$ & $\tau$ & $+\sigma_{\tau}$ & $-\sigma_{\tau}$ & N(o-NH$_2$D)        & $+\sigma_{N}$      & $-\sigma_{N}$     & N(p-NH$_3$)           & R$_D$           & max\{R$_D$\} & min\{R$_D$\} \\
     &                  &             &             & K        & K                  & K                  &        &                  &                  & 10$^{13}$ cm$^{-2}$ & 10$^{13}$ cm$^{-2}$ & 10$^{13}$ cm$^{-2}$ & \multicolumn{2}{l}{10$^{13}$ cm$^{-2}$} &              &              \\

\hline\hline
C01  & 0.33             & 0.04        & 0.04        & 5.7      & -                  & -                  & 0.1    & -                & -                & $<$ 0.7             & -                   & -                  & 6.1 $\pm$ 1.9         & $<$ 0.11        & -            & -            \\
C02  & 0.27             & 0.02        & 0.01        & 6.6      & 1.3                & 0.9                & 0.5    & 0.2              & 0.2              & 1.5                 & 1.1                 & 0.6                & 15.1 $\pm$ 3.8        & 0.10            & 0.23         & 0.05         \\
C03  & 0.25             & 0.02        & 0.00        & 7.1      & 0.9                & 1.0                & 1.6    & 0.2              & 0.2              & 6.7                 & 1.9                 & 1.2                & 24.3 $\pm$ 2.2        & 0.27            & 0.39         & 0.20         \\
C04  & 0.35             & 0.05        & 0.02        & 5.7      & -                  & -                  & 0.1    & -                & -                & $<$ 0.5             & -                   & -                  & 5.8 $\pm$ 1.5         & $<$ 0.09        & -            & -            \\
C06  & 0.38*            & 0.12        & 0.12        & 6.6      & 1.3                & 0.9                & 0.2    & 0.3              & 0.1              & 1.9                 & 2.1                 & 1.0                & 22.1 $\pm$ 2.8        & 0.09            & 0.21         & 0.03         \\
C07  & 0.66             & 0.02        & 0.01        & 5.6      & -                  & -                  & 0.1    & -                & -                & $<$ 0.3             & -                   & -                  & 8.2 $\pm$ 1.6         & $<$ 0.04        & -            & -            \\
C08  & 0.30             & 0.01        & 0.01        & 6.6      & 1.3                & 0.9                & 0.4    & 0.2              & 0.2              & 1.4                 & 1.1                 & 0.6                & 24.7 $\pm$ 9.6        & 0.06            & 0.17         & 0.03         \\
C09  & 0.38*            & 0.12        & 0.12        & 5.7      & -                  & -                  & 0.1    & -                & -                & $<$ 0.6             & -                   & -                  & 4.7 $\pm$ 1.2         & $<$ 0.12        & -            & -            \\
C10  & 0.52             & 0.01        & 0.01        & 6.6      & 1.3                & 0.9                & 0.6    & 0.4              & 0.2              & 1.6                 & 1.1                 & 0.6                & 14.4 $\pm$ 5.8        & 0.11            & 0.31         & 0.05         \\
C11  & 0.41             & 0.01        & 0.01        & 6.6      & 1.3                & 0.9                & 0.4    & 0.3              & 0.1              & 1.7                 & 1.2                 & 0.6                & 22.0 $\pm$ 7.0        & 0.08            & 0.19         & 0.04         \\
C12  & 0.29             & 0.01           & 0.01           & 6.7      & 0.9                & 0.7                & 1.1    & 0.2              & 0.2              & 7.5                 & 2.2                 & 1.7                & 30.7 $\pm$ 2.8        & 0.24            & 0.35         & 0.17         \\
C13  & 0.38*            & 0.12        & 0.12        & 6.6      & 1.3                & 0.9                & 0.5    & 0.5              & 0.3              & 1.9                 & 2.2                 & 1.0                & 21.9 $\pm$ 2.4        & 0.09            & 0.21         & 0.04         \\
C14  & 0.56             & 0.01        & 0.01        & 6.6      & 1.3                & 0.9                & 0.6    & 0.3              & 0.2              & 2.0                 & 1.4                 & 0.7                & 23.4 $\pm$ 7.8        & 0.08            & 0.21         & 0.04         \\
C15  & 0.22             & 0.01        & 0.06        & 5.7      & -                  & -                  & 0.2    & -                & -                & $<$ 0.9             & -                   & -                  & 6.2 $\pm$ 1.7         & $<$ 0.14        & -            & -            \\
C16  & 0.31             & 0.01        & 0.01        & 6.2      & 0.6                  & 0.5                & 1.9    & 0.1              & -                & 11.2                & 2.6                 & 1.2                & 31.9 $\pm$ 4.9        & 0.35            & 0.51         & 0.27         \\
C17  & 0.27             & 0.07        & 0.09        & 5.7      & -                  & -                  & 0.2    & -                & -                & $<$ 0.9             & -                   & -                  & 5.1 $\pm$ 1.8         & $<$ 0.18        & -            & -            \\
C18  & 0.41             & 0.04        & 0.01        & 6.6      & 1.3                & 0.9                & 0.3    & 0.3              & 0.1              & 1.6                 & 1.2                 & 0.6                & 11.7 $\pm$ 2.9        & 0.13            & 0.32         & 0.07         \\
C19  & 0.45             & 0.01        & 0.01        & 6.6      & 1.3                & 0.9                & 0.6    & 0.4              & 0.2              & 1.4                 & 1.0                 & 0.5                & 12.9 $\pm$ 4.5        & 0.11            & 0.29         & 0.05         \\
C20  & 0.38*            & 0.12        & 0.12        & 5.7      & -                  & -                  & 0.1    & -                & -                & $<$ 0.6             & -                   & -                  & 5.0 $\pm$ 1.2         & $<$ 0.12        & -            & -            \\
C21  & 0.42             & 0.10        & 0.09        & 5.7      & -                  & -                  & 0.1    & -                & -                & $<$ 0.5             & -                   & -                  & 4.2 $\pm$ 1.0         & $<$ 0.13        & -            & -            \\
C22  & 0.47             & 0.11        & 0.05        & 5.7      & -                  & -                  & 0.1    & -                & -                & $<$ 0.5             & -                   & -                  & 3.6 $\pm$ 1.6         & $<$ 0.13        & -            & -           
\end{tabular}
\end{table}
\end{landscape}

%%%

\section{Results} 
\label{sec:results}

Out of twenty-two sources, o-NH$_2$D was detected in thirteen cores. The physical locations of detections and non-detections are depicted in Figure \ref{fig:Cepheus}. The spread of non-detections vs detections is fairly uniform throughout the L1251 structure. However, the cores with detections tend to lie in regions with a higher H$_2$ column density. 

We now compare the physical properties of cores detected in o-NH$_2$D and cores not detected in o-NH$_2$D.  We also analyze how the deuterium fraction varies with various core properties.  Lastly, we discuss the implications of the results for core evolution.

\subsection{Abundance \& Deuterium Ratio}

Comparing our column densities with the \textit{Herschel}-derived H$_2$ column density map, we find the average abundance of deuterated ammonia, [o-NH$_2$D] $= 1.3^{+0.8}_{-0.5} \times 10^{-9}$ for the detected subset  (H$_2$ column density from \citealt{2017ApJ...850....3K}). 
The average $3$-sigma upper limit on abundance for the non-detected subset is $0.7 \times 10^{-9}$.

Using column densities of p-NH$_3$ found by \citealt{2017ApJ...850....3K}, the median deuteration ratio R$_D$, or $\frac{[o-NH_2D]}{[p-NH_3]}$, is $0.11$ for all of the sources including the 3 $\sigma$ upper limits. The median R$_D$ for the subset of sources that are detections is 0.10, and the median 3$\sigma$ upper limit to R$_D$ for non-detections is 0.12. 
We did not correct the published p-NH$_3$ column densities for the filling fraction in the GBT beam since the column density is not sensitive to filling fraction when it is near $1$ (see \ref{section:fillingfraction}).
We note that the error bars for R$_D$ are not traditional $1\sigma$ errorbars but instead show the range of R$_D$ values given a $T_{ex}$ range of $5.7$ K and $8.0$ K.
They show the range of systematic uncertainty in R$_D$ and are not statistical uncertainties.

\begin{figure}
    \centering
    \includegraphics[width=85mm]{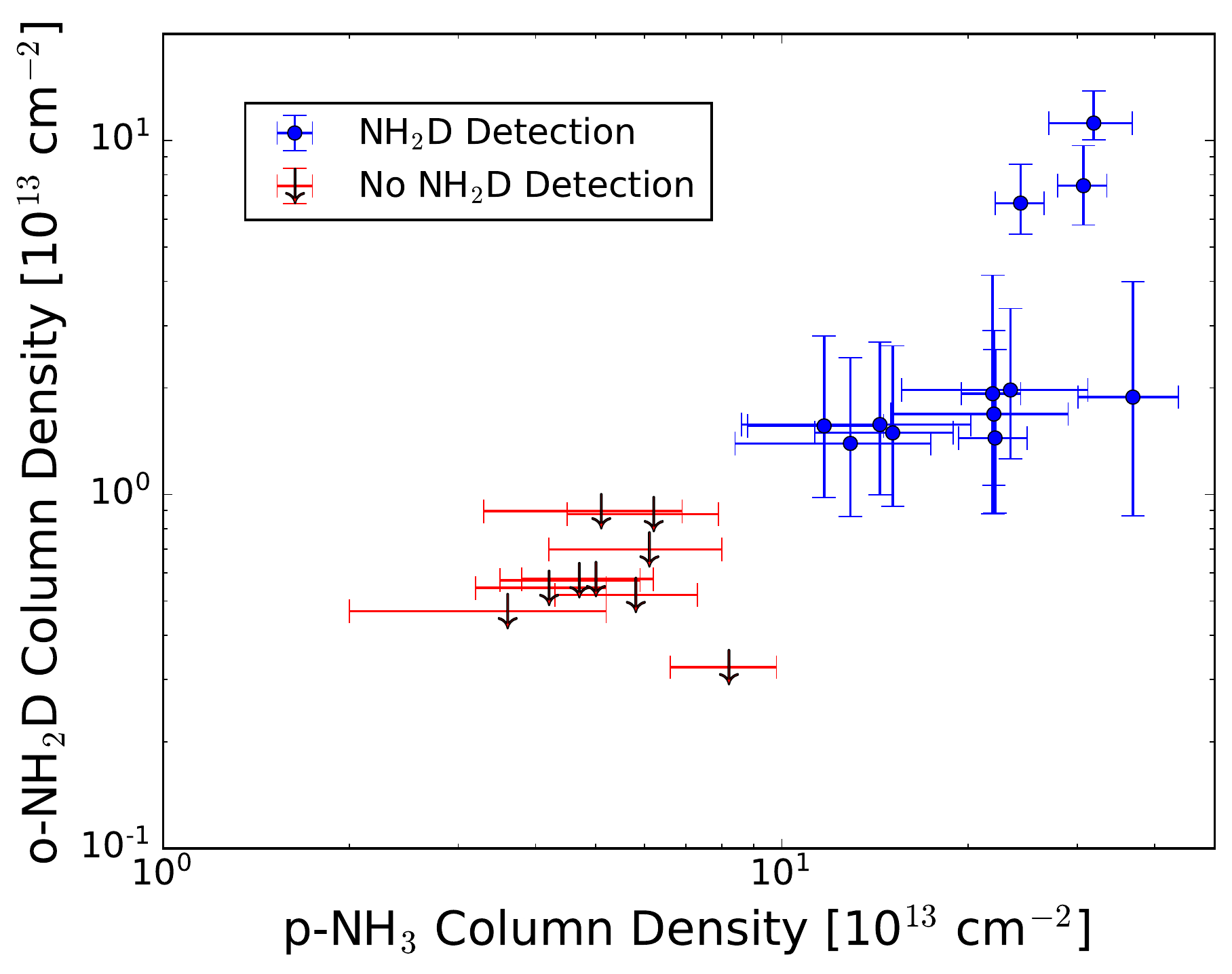}
    %\centering
    \caption{The column density of o-NH$_2$D in blue plotted vs. the column density of p-NH$_3$. A column density threshold in p-NH$_3$ of 10$^{14}$  cm$^{-2}$ for detection of o-NH$_2$D is visible; this is due to observational sensitivity effects.}
    \label{fig:cdvscd}
\end{figure}

Figure \ref{fig:cdvscd} shows a comparison of column densities for o-NH$_2$D and p-NH$_3$. 
One striking feature of Figure \ref{fig:cdvscd} is that there is a column density threshold in p-NH$_3$ of $10^{14}$ cm$^{-2}$ for detection of o-NH$_2$D, as all of the o-NH$_2$D detections have p-NH$_3$ column densities above this threshold.
Our observing strategy targeted all of the NH$_3$-identifed  cores in L1251 to a similar spectrum noise level.
Since the sources not detected in o-NH$_2$D are also the sources with the lowest column density of p-NH$_3$, then the constraints on $R_D$ are not as good, at a given spectral noise level, compared to a source with a higher p-NH$_3$ column density.
This results in the median 3-sigma upper limit of $R_D$ for non-detections to be comparable to the median $R_D$ for detections.
While an observing strategy that integrates for different amounts of time on each source to achieve a similar sensitivity to $R_D$ for all sources would be desirable (i.e. $3\sigma_{R_D}$ upper limit $= 0.05$), it would require a significantly longer investment (additional multi-$100$ hours) in observing time that was not practical for this current survey.
 
% Figure fig, ax = plt.subplots()

\subsection{Comparisons with Mass and Radius}

\begin{figure}
%\plotone{cdvsmassradius.png}
    \centering
%    \includegraphics[width=78mm]{massvsradius.png}
%\begin{multicols}{2}
    \includegraphics[width=85mm]{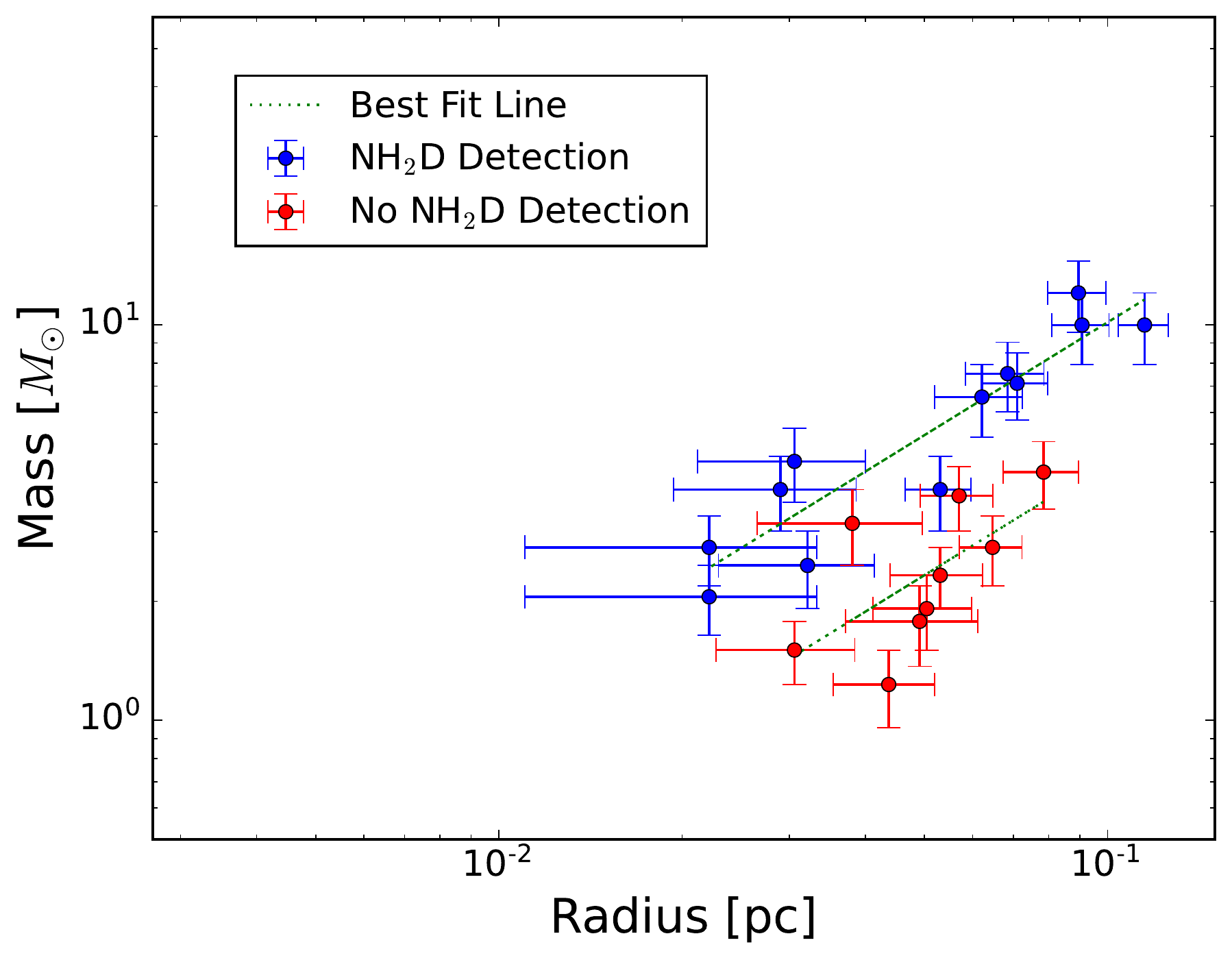}
    \includegraphics[width=85mm]{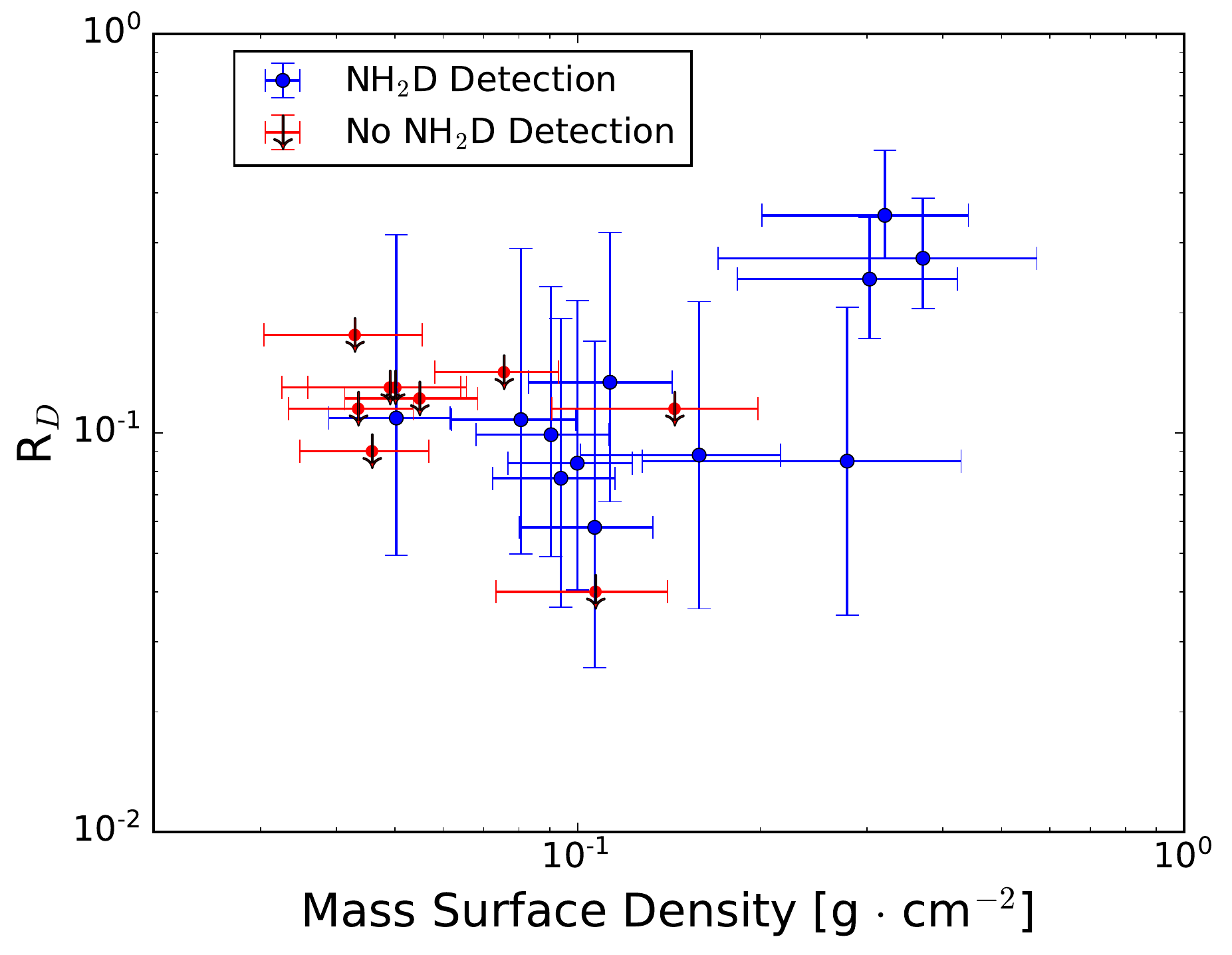}
%    \end{multicols}{2}
    \caption{Top: Mass vs. radius for all cores with best fit lines for cores with o-NH$_2$D detections (blue) and o-NH$_2$D non-detections (red). Cores with a o-NH$_2$D detection (p-NH$_3$ column density $> 10^{14}$ cm$^{-2}$) have a Kendall $\tau$ and respective p-value of 0.76 and 0.0007, while those without a detection (p-NH$_3$ column density $< 10^{14}$ cm$^{-2}$) have a Kendall $\tau$ and respective p-value of 0.61 and 0.025. Bottom: Mass surface density vs. R$_D$.  Errors in R$_D$ are not $1\sigma$ statistical errors but show the range of R$_D$ values given a plausible range in assumed T$_{ex}$ values.}
    \label{fig:massrad}
\end{figure}

The top panel in Figure \ref{fig:massrad} plots mass versus radius for cores with o-NH$_2$D detections and for cores without detections. 
There is clear separation between cores with o-NH$_2$D detection and those without o-NH$_2$D detection.
As we have shown above, the subset of detections and subset of non-detections also correspond to sources with p-NH$_3$ column densities above $10^{14}$ cm$^{-2}$ and below $10^{14}$ cm$^{-2}$ respectively.
It is interesting that cores in the mass radius plot appear to separate based on this arbitrary threshold in p-NH$_3$ column density.
There is the appearance of two mass radius relations.
Relating mass and radius, cores with an o-NH$_2$D detection (p-NH$_3$ column density $> 10^{14}$ cm$^{-2}$) have a Kendall $\tau$ of 0.76 with a p-value of 0.0007. Those without a detection (p-NH$_3$ column density $< 10^{14}$ cm$^{-2}$) have a Kendall $\tau$ of 0.61 and a p-value of 0.025. 
Using a least squares regression, we calculate a best fit line to find the coefficient q in M $\propto$ R$^q$. For cores with o-NH$_2$D detections, q was found to be 0.85. 
For cores without o-NH$_2$D detections, q was found to be 0.93.  
Our observed mass-radius relationship is 
shallower than other dendrogram-based analyses and turbulence-based analyses of the mass-radius relationship in nearby clouds \citep{2010ApJ...716..433K, 2015ApJ...805..185S, 2019ApJ...877...93C}.

The deuterium ratio is plotted against the mass surface density, which depends on both mass and radius, in Figure \ref{fig:massrad}.
There is no clear trend between the deuterium ratios and the mass surface density.
Because of the relative insensitivity to $R_D$ for the low p-NH$_3$ column density sources, there are o-NH$_2$D non-detections with low mass surface densities that have large $3$-sigma upper limits on $R_D$.
Even if a significant (multi-100 hour) observing campaign were undertaken to attempt to detect o-NH$_2$D in those sources, or to lower their upper limits below $R_D < 0.1$, it would not likely change the lack of correlation.

\subsection{Comparisons with Evolutionary Indicators} 

We now explore how the observed deuterium ratio varies with core evolution indicators and interpret the results by comparing to chemical models.

\subsubsection{Core Density}

The central density of the core is an evolutionary indicator as it is expected to increase steadily over time as a prestellar core collapses towards forming a first hydrostatic core.
Since radiative transfer modeling of the dust emission from the cores requires higher angular resolution single-dish submillimeter continuum observations than those currently available for the L1251 region, we are unable to constrain the density profiles of the cores and must therefore calculate the density from average properties (see Equation \ref{eq:avgdensity}).

\begin{figure}
    \centering
    \includegraphics[width=85mm]{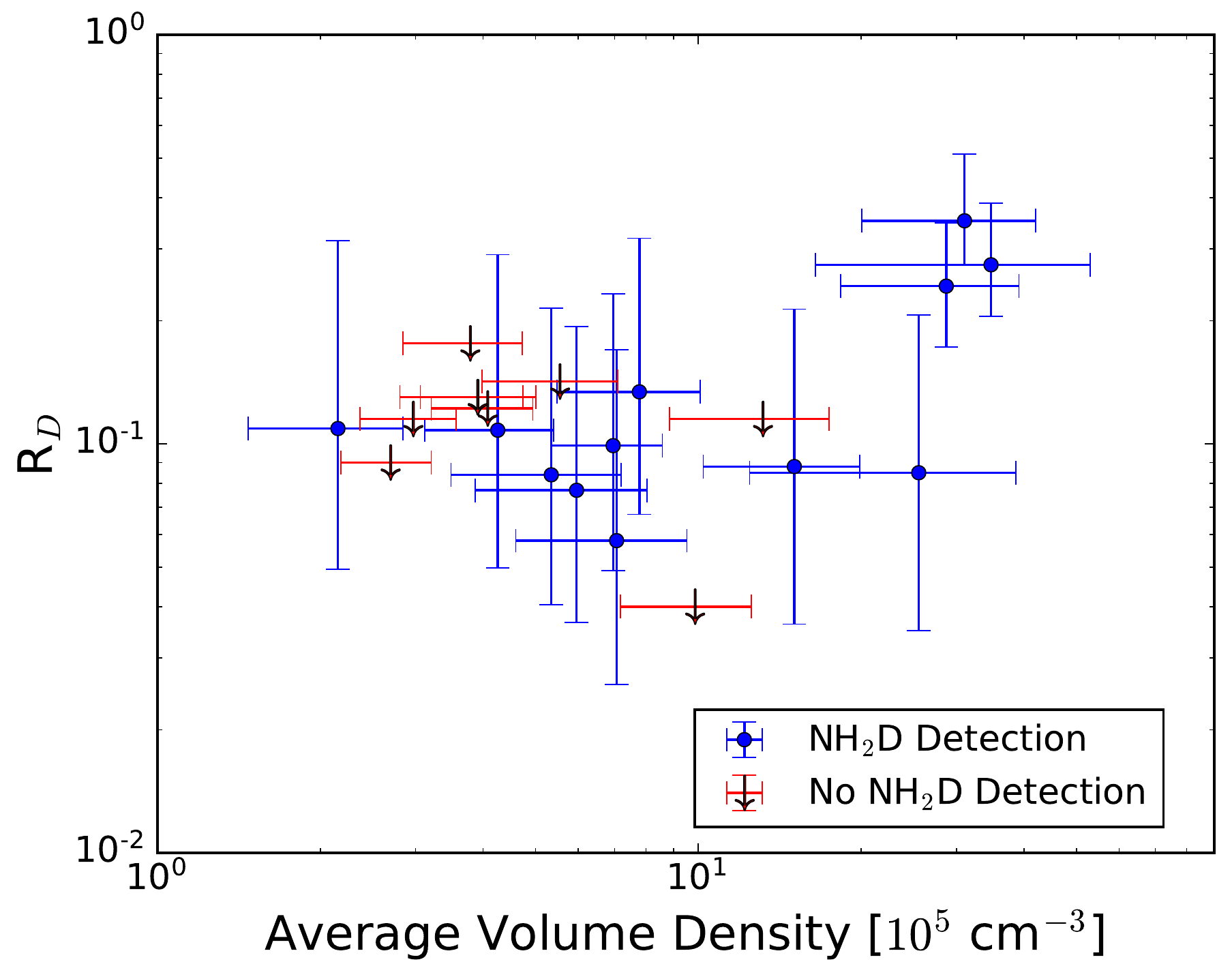}
    \caption{R$_D$, the ratio of o-NH$_2$D column density to p-NH$_3$ column density, plotted against the volume density for cores with and without o-NH$_2$D detection. Errors in R$_D$ are not $1\sigma$ statistical errors but show the range of R$_D$ values given a plausible range in assumed T$_{ex}$ values. The Kendall $\tau$ and p-value (which excludes the non-detections) for this plot are 0.30 and 0.20.}
    \label{fig:RdSphere}
\end{figure}

Figure \ref{fig:RdSphere} plots the deuterium fraction, R$_D$, against the spherically averaged volume density for cores with and without o-NH$_2$D detections. 
The Kendall $\tau$ correlation coefficient and p-value (which excludes the non-detections) for Figure \ref{fig:RdSphere} are 0.30 and 0.20, which indicates no trend between R$_D$ and the average volume density.
The separation between o-NH$_2$D-detected and non-detected cores with the average volume density is not as stark as the separation seen for the mass surface density.
The median average volume density for cores with o-NH$_2$D detection is $7.4 \times 10^4$ cm$^{-3}$, and the median average volume density for cores without o-NH$_2$D detection is $5.4 \times 10^4$ cm$^{-3}$.

\subsubsection{Virial Stability}

A second evolutionary indicator is the virial stability of the cores.
In the Eulerian form of virial equilibrium ($\ddot{I}_{\rm{mom}} = 0$ where $I_{\rm{mom}}$ is the core moment of inertia), twice the kinetic energy ($\Omega_K$) plus the magnetic energy ($\Omega_B$) is in balance with the sum of the gravitational potential energy ($\Omega_G$), twice the energy due to external pressure ($\Omega_P$), and the rate of change of the momentum flux across the core surface \citep{1992ApJ...399..551M,2017stfo.book.....K}. This can be written generally as
\begin{equation}
\label{eq:SimpleVirial}
    2\Omega_K + \Omega_B = \Omega_G + \Omega_P + \frac{d (\textrm{Momentum Flux})}{dt} \;.
\end{equation}
This equation can be expanded into the individual energy terms, with some simplifying assumptions \citep{2017stfo.book.....K}, as 
\begin{equation}
\label{eq:mag}
    3M\sigma^2 + \frac{(B^2 - B_0^2)R^3}{6} = a_1 \frac{GM^2}{R} + 4{\pi}P_{o}{R^3} + \frac{1}{2}\frac{d}{dt}\int_{\partial V} \mu m_H R^2 n_o \vec{v} \cdot d\vec{S} \;.
\end{equation}
The kinetic energy term ($2\Omega_K = 3 M\sigma^2 $) assumes no significant velocity gradients across the core (see \citealt{2021arXiv210805367S}).
We use the mass of the cores, M, tabulated in \citealt{2017ApJ...850....3K} with the velocity dispersion, $\sigma$, calculated in \S3.1.
$B$ and $B_o$ are the uniform magnetic field interior and exterior to the core such that this term is the difference in magnetic energy inside and outside the core;  
we explore this term in more detail below.
The radius of the core, R, is the dendrogram derived source size (calculation discussed in Section \ref{section:fillingfraction}).
$a_1$ is a constant correction factor for the density profile of the core.  
For a uniform density sphere, $a_1 = 0.6$, while for a critical Bonnor-Ebert sphere, $a_1 = 0.72$ \citep{2011A&A...535A..49S}.
Since there is only a small variation between these values, we assume a uniform density sphere for all virial calculations. 
$P_{o}$ is the external pressure which we treat as a turbulent ram pressure given by
\begin{equation}
    P_{o} = \mu_p m_{H} n_o \sigma_{o}^2 \; ,
\end{equation}
where $n_o$ is the density at the core boundary and $\sigma_{o}$ is the external velocity dispersion, calculated by averaging the NH$_3$ (1,1) velocity dispersion from the pixels immediately outside the dendrogram-derived core masks.
The final term depends on the time derivative of the surface integral of the dot product of the velocity of gas flowing across the core boundary with the local surface normal vector.
There are no observational constraints on this quantity, but given that it depends on the rate of change of the flow in time, it is likely to be small compared to other terms, especially for cores likely to form low-mass stars.

\begin{figure}
    \centering
    \includegraphics[width=90mm]{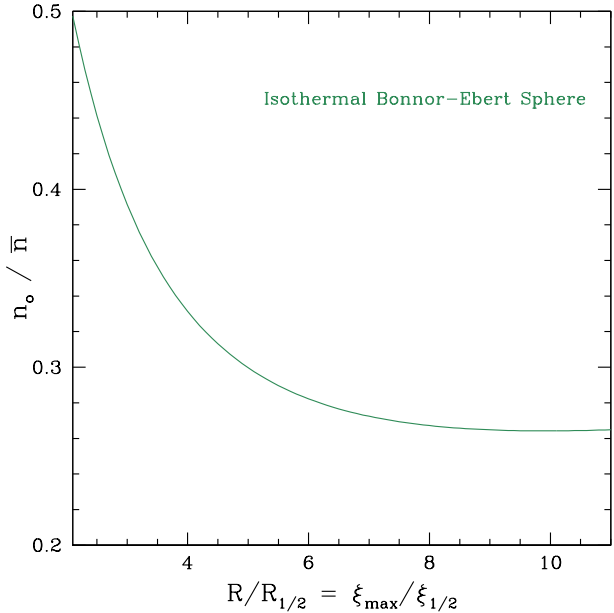}
    \vspace{-1.5cm}
    \caption{The ratio of outer density to average density vs. the ratio of outer radius normalized by the HWHM radius for an isothermal Bonnor-Ebert sphere.  The x-axis spans the range observed for starless cores.}
    \label{fig:BES}
\end{figure}

We need an estimate of the outer density, $n_0$, of the core in order to calculate the external pressure virial term.
The outer number density may be estimated directly from the core average density assuming that the core's density profile is described by a isothermal Bonnor-Ebert sphere (\citealt{1955ZA.....37..217E, 1956MNRAS.116..351B}; see Appendix \ref{section:Appendix}). 
The ratio of outer density to spherically-averaged density ($n_o/\bar{n}$) is a slowly varying function over a wide range of $R/R_{1/2}$ (Figure \ref{fig:BES}),
where $R_{1/2}$ is defined as the radius for which the core density is half of the central density.
The ratio of $n_o/\bar{n}$ not sensitive to the ratio $R/R_{1/2}$ unless the ratio becomes small.
$R_{1/2}$ cannot be determined without radiative transfer modeling of the core; however,
near-infrared observations of isolated starless cores from \cite{Kandori_2005} have constrained 
%$\xi_{max}$ to the range $4.7 - 25$ which corresponds to a range in 
$R/R_{1/2} = 2.1 - 11$ (N.B. $R/R_{1/2} = \xi_{max}/\xi_{1/2}$ where $\xi_{max}$ is the maximum dimensionless radius and $\xi_{1/2} = 2.274$ is the dimensionless radius where the density falls to half of the peak density; \citealt{1939isss.book, 1949ApJ...109..551C}).
Across this range , there is a less than factor of 2 difference in the ratio of the outer volume density to the average volume density.
While the near-infrared observations targeted isolated cores and the cores in our survey are clustered within filamentary structure, this insensitivity to the exact choice of $R/R_{1/2}$ does not introduce more than a factor of 2 uncertainty in our estimates of the outer density.
We use the median value and range in $R/R_{1/2}$ $= 3.6^{+7.4}_{-1.5}$ from starless and prestellar cores in \cite{Kandori_2005} to calculate $n_o$ using the curve shown in Figure \ref{fig:BES}.
This corresponds to multiplying the average density by a factor of $0.35^{+0.14}_{-0.09}$.

\begin{figure}
    \centering
    \includegraphics[width=85mm]{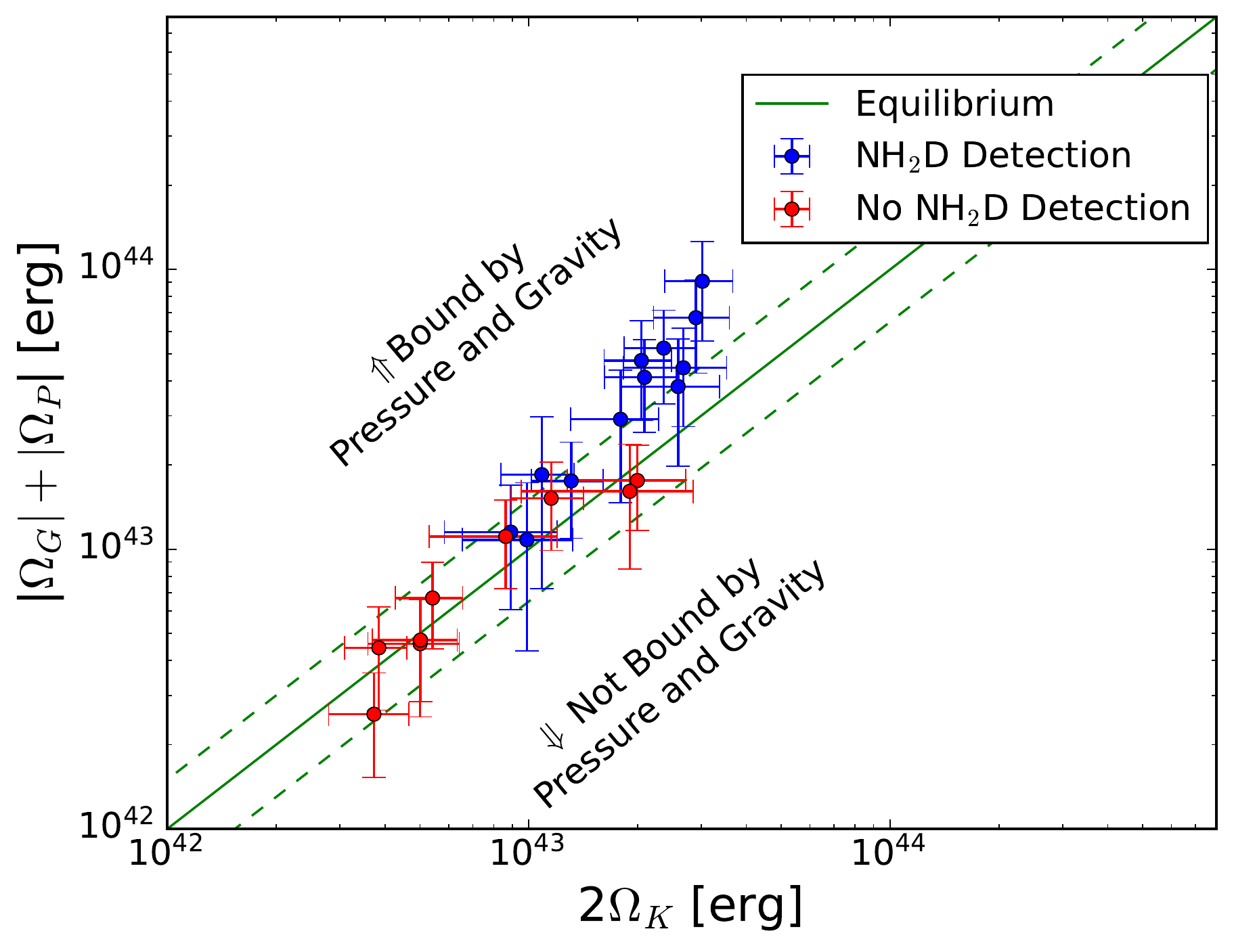}
    \caption{Kinetic energy ($\Omega_K$) compared to the sum of gravitational potential energy and external gas pressure ($\Omega_G + \Omega_P$).  The green solid line corresponds to virial equilibrium (in the $B_{eff} = 0$ limit) while the dashed lines correspond to factors of 2 uncertainty in virial equilibrium that may be due to density profile or geometrical effects.}
%    \label{fig:virialplot}
\label{fig:fourvirialplot}
\end{figure}

The ratio of $|2\Omega_K/\Omega_G|$ is often referred to as the virial parameter, $\alpha$; however, the terms due to external pressure and the magnetic field should not be ignored.
A more accurate virial parameter should be defined as $\alpha = (2\Omega_K + \Omega_B)/(|\Omega_G| + |\Omega_P|)$ where a ratio $< 1$ would indicate that the core is bound by external pressure and gravity.  In practice, the uncertainty in determining the magnetic field strength makes it difficult to calculate an accurate $\Omega_B$ term.
The relative importance of these virial energy terms, excluding the magnetic energy, are explored in Figure \ref{fig:fourvirialplot}.
The terms on the y-axis are attempting to collapse the core while the term on the x-axis is providing internal support.
There is separation seen between o-NH$_2$D-detected cores (p-NH$_3$ column density $> 10^{14}$ cm$^{-2}$) and o-NH$_2$D non-detected cores (p-NH$_3$ column density $< 10^{14}$ cm$^{-2}$) with the majority of cores with o-NH$_2$D detections tending to have larger virial energy terms than cores that are o-NH$_2$D non-detections by about a factor of 2.
It appears that most of the cores are close to virial equilibrium when we account for errorbar overlap.
50\% of cores have errorbars that overlap with being bounded by external pressure plus gravity. The errorbars of 90\% of the cores overlap with the dashed lines which represent virial equilibrium. A lesser amount (31\%) of cores have errorbars that overlap with not being bound by pressure and gravity.
Only one core has virial parameters that suggest it is not likely near virial equilibrium.

% New Figure 10

\begin{figure}
    \centering
    \includegraphics[width=85mm]{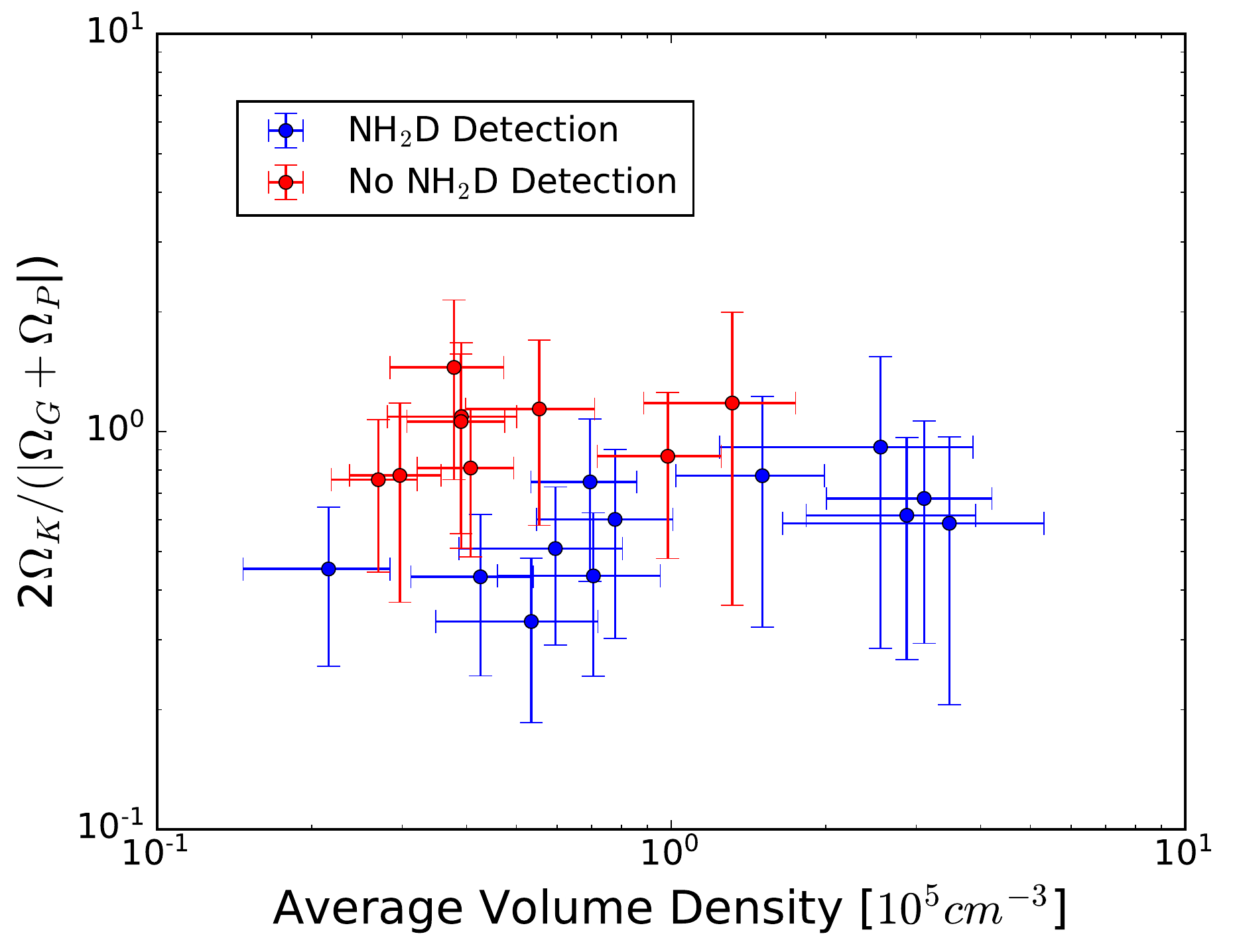}
    \caption{Virial parameter excluding the magnetic term plotted against average density. We assume the cores are in equilibrium ($d^2I_{\rm{mom}}/dt^2 = 0$). Both the virial parameter and average density are possible evolutionary indicators. Although cores with non-detections tend to have a higher virial parameter, the Kendall $\tau$ coefficients and p-values do not indicate correlation. The Kendall $\tau$ correlation and p-value for detections is 0.42 and 0.062, while for non-detections those values are 0.38 and 0.18.}
    \label{fig:virialplot}
\end{figure}

The virial parameter, excluding the magnetic field term, is compared with the core density in Figure \ref{fig:virialplot}. 

It is interesting to note that these two purported evolutionary variables do not correlate well with each other.
The sources with the highest average volume density, while all detected in o-NH$_2$D, are only near the median range in virial parameters.
This result illustrates that these indicators are not simple monotonic evolutionary indicators of core evolution.
One important caveat is that the virial analysis assumes that the cores are in equilibrium ($d^2I_{\rm{mom}}/dt^2 = 0$) which may not be true.

\begin{figure}
    \centering
    \includegraphics[width=85mm]{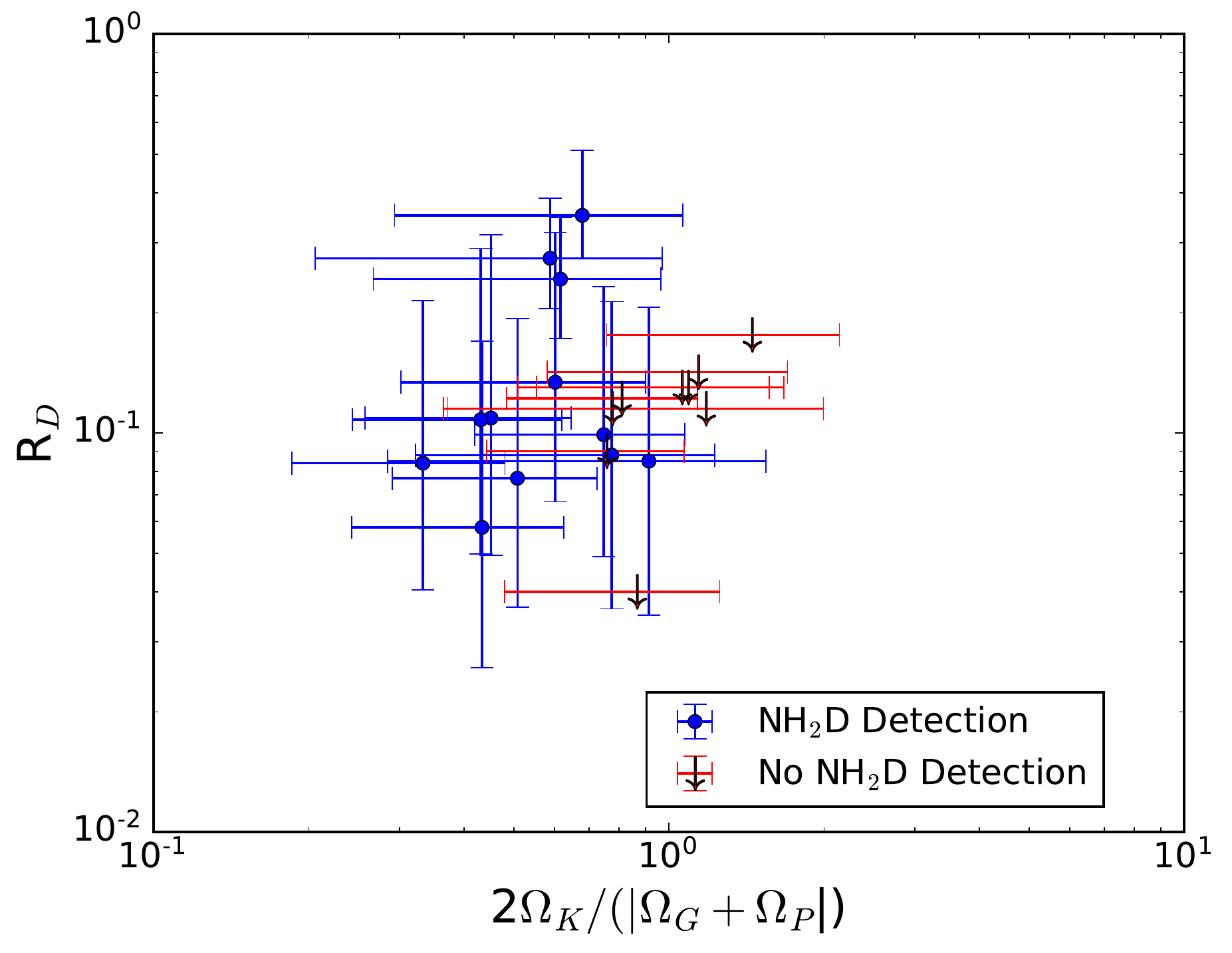}
    \caption{R$_D$ plotted against the virial parameter including gravity and pressure but excluding the magnetic field term. Errors in R$_D$ are not $1\sigma$ statistical errors but show the range of R$_D$ values given a plausible range in assumed T$_{ex}$ values. Again, there is not significant correlation between these parameters: detections have a Kendall $\tau$ of 0.15 (with a p-value of 0.54) and non-detections have a Kendall $\tau$ of 0.57 (with a p-value of 0.035).}
    \label{fig:rdvirial}
\end{figure}

Figure \ref{fig:rdvirial} plots the deuterium fraction, R$_D$, against the virial parameter, excluding the magnetic field term. 
This plot demonstrates that there is not significant correlation between ammonia deuterium fractionation and the virial parameter.
This lack of correlation is very similar to that seen between $R_D$ and the mass surface density and the average volume density.

% Figure 

\subsubsection{A Caveat About Magnetic Fields}

\begin{figure}
    \centering
    \includegraphics[width=85mm]{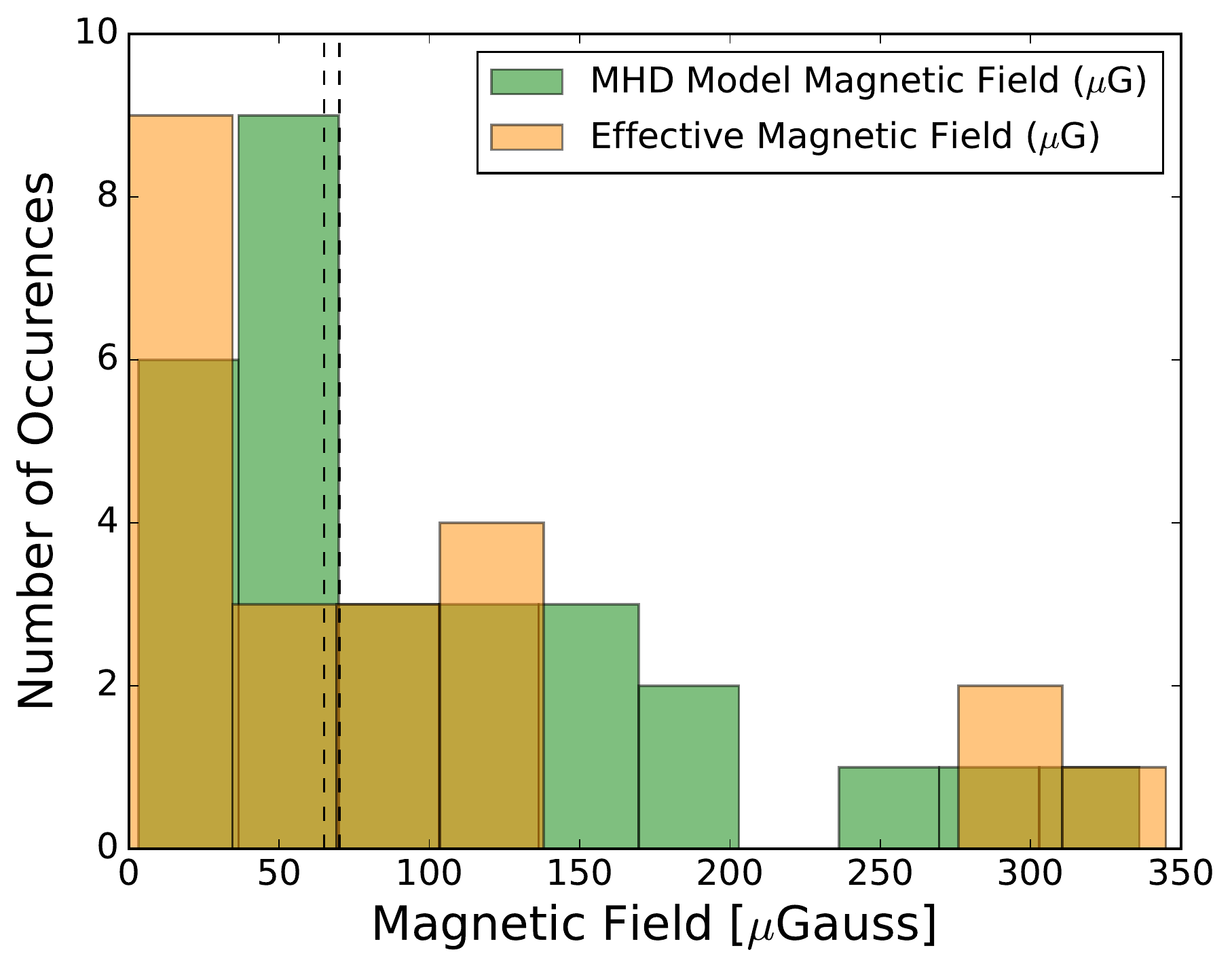}
    \caption{In orange: Histogram of the effective magnetic fields ($\sqrt{B^2 - B_0^2}$) required for equilibrium  calculated by balancing virial parameters 2$\Omega_K$, $\Omega_G$, $\Omega_P$, and $\Omega_B$ and using Equation \ref{eq:mag}. In green: histogram of effective magnetic fields determined from a MHD simulation of dense cores with comparable mass to those observed in L1251. Black dotted vertical lines show the median of both data sets.}
    \label{fig:magfieldhistogram}
\end{figure}

The effective magnetic field is defined as
\begin{equation}
B_{\rm{eff}} = \sqrt{B^2 - B_o^2} \;,
\end{equation}
and is proportional to the square root of the difference in magnetic energy density inside the core and outside the core that is needed for virial equilibrium (Equation \ref{eq:mag}). 
Because of a lack of magnetic field measurements for this region, we cannot calculate this term from observations taken toward the cores in L1251. 
Instead, we reverse the problem to calculate an estimate of the effective magnetic field strength needed to balance the remaining three virial terms. 
Figure \ref{fig:magfieldhistogram} depicts a histogram of the effective magnetic field strength found using this method (in orange). 
We find a median effective B-field strength of 65 $\mu$G.

Magnetic field strength in dense star-forming regions is notoriously hard to constrain because of limited accuracy in the observational techniques \citep{2012ARA&A..50...29C}.
We need spatially resolved estimates of the magnetic field inside and outside of cores to estimate the importance of the magnetic virial term.
The measurements of the B field strength in dense starless and prestellar cores mostly comes from analysis of the dispersion of polarization directions from submillimeter dust continuum polarimetry \citep{2004ApJ...600..279C, 2019ApJ...877...43L, 2019ApJ...877...88C, myers2021magnetic, 2022arXiv220509134C}.
Typical values for the total magnetic field strength (corrected from the plane-of-the-sky component that is measured) span from $100 - 200$ $\mu$G with large uncertainties in low-mass star forming regions that are similar to L1251.
Observations of the magnetic field strength in the surrounding cloud come from both Zeeman measurements of low density gas tracers and from the dispersion of optical and near-infrared polarization measurements of background starlight.
For instance, in the Taurus and Ophiuchus molecular clouds, \citealt{Troland_1996} found a magnetic field of $\sim 10$ $\mu$G using OH Zeeman effect measurements.
Polarization observations of background stars in the optical and near-infrared of the central Taurus molecular cloud find magnetic fields in the range of 5 $\mu$G to 82 $\mu$G for A$_{\rm{V}} < 9$ mag \citep{2011ApJ...741...21C}.
The large spread in observed magnetic field estimates and low accuracy in those estimates, both inside and outside of dense cores, makes it very difficult to pin down an accurate contribution from the magnetic energy term in the virial equation.

Since the observations do not provide reliable constraints, we compare to theoretical simulations of core evolution.
\citealt{2020MNRAS.497.4517S} analyzed core evolution in a 3-dimensional, adaptive mesh refinement, ideal magnetohydrodynamic (MHD) simulation using the ORION2 code \citep{1999JCoAM.109..123K,2007ApJ...667..626K,2012ApJ...745..139L}. 
The initial conditions were chosen to be consistent with a typical molecular cloud in the Gould's Belt: isothermal at $10$ K and a mass of 3762 M$_{\odot}$.% (taken from run W2T2 of \citealt{2015ApJ...811..146O}) on a 256$^3$ grid.
%The initial magnetic field has a uniform direction and a strength of $13$ $\mu$G with an initial thermal pressure to magnetic pressure ratio of $0.1$.
%A barytropic equation of state is adopted with a critical density above which the gas changes from being isothermal to adiabatic (see section 2.2 of \citealt{2020MNRAS.497.4517S} for details).
%After an initial turbulent injection phase where a turbulent steady state is achieved with power-law $P(k) \sim k^{-2}$, self-gravity is turned on and the simulation is followed for 70\% of the global free-fall time (1.5 Myrs). 
Dendrograms based on density contrast of a factor of $\geq 3$ above the threshold density of $1.8 \times 10^4$ cm$^{-3}$ were analyzed to follow the core properties during the simulation evolution \citep{2020MNRAS.497.4517S}.
We have re-analyzed those dendrograms for the subset of cores with a similar mass range to those observed in L1251 at 40\% of the global cloud free-fall time (near the midpoint of the simulation; t$_{ff}$ = 1.5 Myr) to derive the histogram of effective magnetic field strengths shown in green in Figure \ref{fig:magfieldhistogram}.
While the distribution is slightly different to what is required for virialization of the observed cores in L1251, the median effective magnetic field strength found by the MHD simulation is $70$ $\mu$G.
We note that some cores in the simulation had surface magnetic energy terms that were larger than the interior magnetic energy. If those sources are also included, the median effective magnetic field drops to $40$ $\mu$G.
The results from this model only represent one set of initial conditions and do not include non-ideal MHD effects, such as ambipolar diffusion, which may change the distribution of theoretical effective magnetic fields.
Nevertheless, this calculation implies that we cannot ignore the magnetic field in the virial parameter, and that it likely provides non-negligible support within the virial equation. 
Variations in the real magnetic virial terms, which are unaccounted for in typical virial analyses, may be a significant source of scatter that can de-correlate the comaprison of the virial parameter with other evolutionary variables.

\subsection{Implications for Core Evolution}

\begin{figure}
    \centering
    \includegraphics[width=85mm]{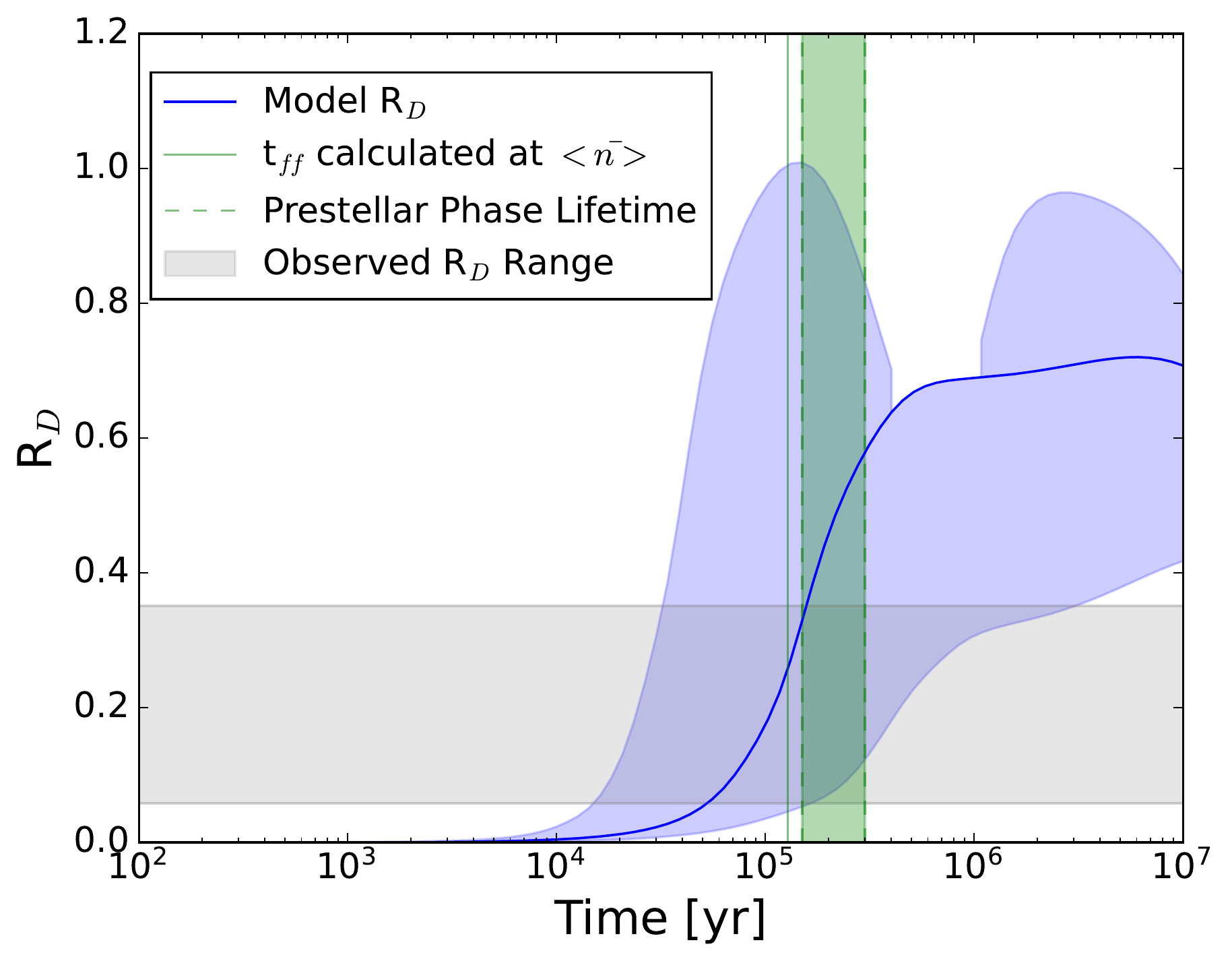}
    %\centering
    \caption{Static, isothermal chemical model prediction of $R_D =$ [o-NH$_2$D]/[p-NH$_3$] run at the median density ($7.4 \times 10^4$ cm$^{-3}$) of cores detected in o-NH$_2$D (solid line). The blue shaded regions represent the range in the model values at our maximum average core density ($3.48 \times 10^5$ cm$^{-3}$) and at our minimum average core density ($2.16 \times 10^4$ cm$^{-3}$). The grey shaded region represents the range in R$_D$ from our calculations, the green line represents the free-fall time calculated at the median average density, $\bar{<n>}$, and the green shaded region with the dotted green line shows the prestellar phase lifetime. The sharp changes in the upper error of R$_D$ is due to a transient change in the ratio of abundances, not the abundances themselves.}
    \label{fig:Rdmodel}
\end{figure}

Static chemical models provide a good starting point for comparison with the observed $R_D$ values since collapsing core chemo-dynamical models depend on which hydrodynamical collapse solution is adopted (i.e. \citealt{2004ApJ...617..360L, 2012ApJ...760...40A, 2018MNRAS.477.4454H}).
The static, isothermal ($10$ K for both dust and gas) prestellar core models of \citealt{2019A&A...631A..63S} follow the spin statistics of the molecules through chemical reactions (i.e. \citealt{1977MolPh..34..477Q, 2004JMoSp.228..635O, 2016JChPh.145g4301S}) and predicts the variation of o-NH$_2$D abundance with time at constant densities ($10^4$, $10^5$, and $10^6$ cm$^{-3}$). 
The median core density of the o-NH$_2$D detected sub-sample is $7.4 \times 10^4$ cm$^{-3}$. 
The \citealt{2019A&A...631A..63S} complete scrambling model, with the same initial conditions and parameters as published models, was re-run for this median volume density and with a typical extinction of the interstellar radiation field of $A_{\rm{V}} = 5$ mag at the core boundary (see Figure \ref{fig:Rdmodel}).
The deuteration ratio for the median core density in our o-NH$_2$D detected sample rises very quickly over a short time period of  $10^5$ years $\pm$ a few $10^4$ years.
The model predicts $R_D$ to rise above $0.1$ on timescales of $7 \times 10^4$ years and to be as high as $R_D = 0.3$ on timescales of $\sim 10^{5}$ years.
This model is heavily dependent on density, 
with higher core densities reaching a given $R_D$ value faster than lower core densities.
For our range of observed densities, this results in an order of magnitude spread in the chemical deuteration timescale.

The deuteration chemical timescale is comparable to the free-fall timescale, $t_{\rm{ff}} = \sqrt{3\pi / 32 G \mu m_H \bar{n}} \sim 10^5$ years, at the observed median core density.
The calculated free-fall timescale is shown as the green line in Figure \ref{fig:Rdmodel} along with our calculated range of R$_D$. 
Prestellar phase lifetimes are typically longer than the free-fall time by factors of $1.5$ to $3$ \citep{Andr__2014}. We show this range as a green shaded region in Figure \ref{fig:Rdmodel}.
The prestellar phase lifetime is defined as the average time that a prestellar core of a given density will remain prestellar.
It is determined from counting the number of prestellar cores at different average densities identified in \textit{Herschel} surveys of entire molecular clouds and then comparing those numbers to the number of objects, usually Class II protostars, with a known phase lifetime (see \citealt{2009ApJS..181..321E}). 
The observed median density of $7.4 \times 10^4$ cm$^{-3}$ corresponds to a prestellar phase-lifetime of $\sim 1.5-3 \times 10^5$ years, based on the prestellar core counting statistics presented in \citealt{Andr__2014}. 
Given the range in $R_D$ that we observe, the deuteration chemical timescale from the \citealt{2019A&A...631A..63S} model is less than, and therefore, not in tension with the prestellar phase lifetime of the subset of cores with o-NH$_2$D detections.

Relevant published chemo-dynamical models of prestellar core evolution are the models of \citealt{2018MNRAS.477.4454H} that calculate the deuteration spin-statistics chemistry for isothermal, collapsing prestellar cores with a one-dimensional Larson-Penston collapse solution \citep{1969MNRAS.145..271L, 1969MNRAS.145..457P, 1972MNRAS.157..121L}. 
This model under-predicts the observed median value of the median observed o-NH$_2$D column density ($1.8 \times 10^{13}$ cm$^{-2}$) at the average densities ($7.4 \times 10^4$ cm$^{-3}$) of cores with detections (see Figure 5 of \citealt{2018MNRAS.477.4454H}).
The model also under-predicts the deuterium fraction with $R_D < 0.1$ for all model core densities that are $< 3 \times 10^{5}$ cm$^{-3}$.
To reconcile these differences, it is important to highlight that the results of all astrochemical models depend on the assumed initial conditions.

The median observed mass of the NH$_3$-identified cores in L1251 is $3.4$ M$_{\odot}$ (spanning a range from $1.2 - 12$ M$_{\odot}$) while the models of \citealt{2018MNRAS.477.4454H} start with a higher mass ($7$ M$_{\odot}$), isothermal ($10$ K), core with density of $10^4$ cm$^{-3}$.
By changing the initial mass, initial density, isothermal temperature, cosmic-ray ionization rate, initial fractional abundances (in the gas or on the grain mantles), or the collapse solution would change the predicted abundances.
Source-specific chemo-dynamical modeling is beyond the scope of this paper as it requires detailed physical models of each source. Our observations of detected cores, however, generally agree within a factor of two with the state-of-the-art published deuteration models of prestellar cores.

A remaining question to answer is: why does $R_D$ not correlate with physical or evolutionary variables?
One very important caveat is that none of the evolutionary variables are perfect indicators of the evolutionary state of a core.
As an example, the central density of a core is not the sole evolutionary parameter that fully describes its evolutionary state because cores at the same central density can have very different chemical evolutionary histories.
A good example of this chemical history dichotomy is L1521E and L1498, two prestellar cores located in the Taurus molecular cloud that have similar central densities (few $\times 10^5$ cm$^{-3}$; \citealt{2002ApJ...569..815T,2004A&A...414L..53T}).
L1521E is a chemically young core (low CO depletion, strongly peaked in emission from carbon chain molecules such as CCS and HC$_5$N, and low deuteration ratios; \citealt{2002ApJ...565..359H, 2019A&A...630A.136N, 2021MNRAS.504.5754S}) while L1498 is a chemically mature core (significant CO depletion, depleted carbon chain molecules in the center, and higher deuteration ratios; \citealt{1998ApJ...507L.171W, 2005ApJ...632..982S, 2006A&A...455..577T, 2013A&A...559A..53M, 2021arXiv210508363J}).
This implies that L1521E has only recently ($\leq 10^5$ years) arrived at its present density while L1498 has spent a longer time at its present density.
Thus, density alone is not a sole evolutionary indicator but instead must be used in combination with other indicators to characterize more fully the evolutionary state of a prestellar core.
Physics-based indicators (e.g., density, virial parameter) need to be used together with chemistry-based indicators (e.g. deuteration, CO depletion factors).

The lack of correlation of $R_D$ with mass surface density, average density, and virial parameter for the L1251 cores ultimately has implications for their rates of evolution.
The significant overlap between o-NH$_2$D detected and non-detected sources at densities below $1.3 \times 10^5$ cm$^{-3}$ and the lack of correlation between $R_D$, $\bar{n}$, and virial parameter in Figures \ref{fig:RdSphere} and \ref{fig:fourvirialplot} are likely indicators of the dichotomy illustrated above for L1521E and L1498, namely, that cores can be observed that have spent different amounts of time at the same density and therefore develop different levels of deuterium fractionation at the same density.
Assuming that the variables with which we have compared $R_D$ values do have some ability, albeit imperfectly, to distinguish between cores in different evolutionary phases, then the lack of correlation in deuteration can be explained by the cores evolving at different rates.
Cores spend different amounts of time evolving through density and therefore develop different levels of deuteration. 
Previous observational evidence points to cores evolving at rates slower that free-fall collapse.
Thus, it is logical that cores would evolve chemically at different rates since any differences in physical mechanisms that mitigates core collapse rates, such as the ambipolar diffusion rate, would result in differential chemical core evolution rates.
A future challenge of dense core studies is to identify which internal or external physical and chemical properties of the cores dominate core evolution rates.

\section{Conclusions}
We observed 22 NH$_3$-identified cores in Cepheus L1251 in the J$_{\rm{K_a K_c}}^{\pm}$ $= 1_{11}^{+} \rightarrow 1_{01}^{-}$ transition of o-NH$_2$D. 
We detected such emission from 13 of the 22 cores with a median baseline rms of $\sigma_{T_{mb}} = 17$ mK. 
The observed hyperfine pattern (including splitting due to the N and D atoms) was fit using CLASS.
We assume a constant excitation temperature (CTEX model) to calculate the o-NH$_2$D column densities for optically thin and optically indeterminate sources. 
Using this data, we draw the following conclusions.

1. All sources detected in o-NH$_2$D have p-NH$_3$ column densities $> 10^{14}$ cm$^{-2}$ while all non-detections have p-NH$_3$ column densities $< 10^{14}$ cm$^{-2}$. The difference between detections and nondetections in this threshold in column density is likely due to the sensitivity limitations of this survey.

2. We observe separation when plotting several quantities against each other between cores with o-NH$_2$D detections and those without which is the same as cores with p-NH$_3$ column densities $>10^{14}$ cm$^{-2}$ and cores with p-NH$_3$ column densities $<10^{14}$ cm$^{-2}$.
The most striking examples are plots of mass vs. radius and virial support (excluding magnetic energy) vs. virial collapse terms.

3. A re-analysis of the virial balance 
finds that the virial parameter (excluding magnetic energy) for 95\% of the cores overlaps within uncertainties with virial equilibrium while the remaining core is bound by external pressure plus gravity. 
However, the magnetic fields in the L1251 cores likely help support the cores against gravitational collapse.
Further study of the magnetic fields in the L1251 dense cores is needed to fully understand their role in core evolution.

4. We find a median deuterium fraction, R$_D$ ([o-NH$_2$D]/[p-NH$_3$]), of 0.11 including the 3 $\sigma$ upper limits for the non-detections.
There are no strong, discernible trends in plots of R$_D$ with any physical or evolutionary parameters.
If the L1251 cores have similar chemical initial conditions, then this result indicates that the cores are evolving through density at different rates.

Observations of the deuteration of ammonia have proven to be helpful chemical evolutionary indicators in these 22 dense cores, although in this study we are limited by the sensitivity-driven selection effects.
Future chemo-dynamical studies that include non-ideal MHD effects can test which physical parameters drive the core evolution rate and also test whether the observed spread in $R_D$ values at different densities can be reproduced.

\section*{Acknowledgements}

We are thankful that we have the opportunity to conduct astronomical research on Iolkam Du'ag and acknowledge the very significant cultural role and reverence that this site has to the Tohono O’odham Nation.
We are grateful to our referee, Erik Rosolowsky, for comments that greatly improved this paper.
We sincerely thank the staff and the operators of the Arizona Radio Observatory (Michael Begam, Kevin Bays, Clayton Kyle, and Robert Thompson) for their assistance with the observations. 
Maria Galloway-Sprietsma was supported by an Arizona NASA Space Grant Consortium internship (PI Swindle; NNX15AJ17H).
Yancy Shirley was supported by National Science Foundation Grant AST-1410190 (PI Shirley).
Olli Sipil{\"a} acknowledges the support of the Max Planck Society.
Rachel Smullen was supported by the U.S. Department of Energy (DOE) Advanced Simulation and Computing (ASC) Program’s Metropolis Fellowship through Los Alamos National Laboratory (LANL).  LANL is operated by Triad National Security, LLC, for the National Nuclear Security Administration (NNSA) of DOE under Contract No. 89233218CNA000001. Calculations in this work were were run upon High Performance Computing (HPC) resources supported by the University of Arizona TRIF, UITS, and Research, Innovation, and Impact (RII) and maintained by the UArizona Research Technologies department.
The 12 m Telescope is operated by the Arizona Radio Observatory (ARO), Steward Observatory, University of Arizona, with the the bulk of operations funding from the State of Arizona and additional support comes from the National Science Foundation (NSF) for instrumentation development. Current NSF support includes Major Research Instrumentation (MRI) funding for Development of a State-of-the-Art Multiband Receiver for Arizona Radio Observatory's New ALMA Antenna (PI Ziurys; AST-1531366), Mid-Scale Innovations Program (MSIP) funding for The Event Horizon Telescope Experiment (PI Marrone; SV5-85009/AST- 1440254), a Faculty Early Career Development (CAREER) program grant for Mapping the Fuel for Star Formation Across Cosmic History (PI Marrone; AST-1653228), an Advanced Technologies and Instrumentation grant for Measuring Reionization and the Growth of Molecular Gas with TIME (PI Marrone; S455656/AST-1910598), and a Partnerships for International Research and Education (PIRE) award for Black Hole Astrophysics in the Era of Distributed Resources and Expertise (PI Psaltis; OISE-1743747).

\section*{Data Availability}

The data underlying this article will be shared on reasonable request to the corresponding authors.

%%%%%%%%%%%%%%%%%%%%%%%%%%%%%%%%%%%%%%%%%%%%%%%%%%



\bibliographystyle{mnras}
\bibliography{nh2d_cepheus} % if your bibtex file is called example.bib

%\begin{thebibliography}{99}
%\end{thebibliography}

%%%%%%%%%%%%%%%%%%%%%%%%%%%%%%%%%%%%%%%%%%%%%%%%%%

%%%%%%%%%%%%%%%%% APPENDICES %%%%%%%%%%%%%%%%%%%%%

\appendix

\section{A Note About the Hyperfine Transitions of ortho-NH$_2$D}
\label{section:AppHyp}

The spectrum of o-NH$_2$D has hyperfine splitting due to electric quadrupole and magnetic dipole interactions between the rotation of the molecule and the spin-1 $^{14}$N and $^{2}$D atoms ($\vec{F}_1 = \vec{J} + \vec{I}_N$ and $\vec{F} = \vec{F}_1 + \vec{I}_D$; \citealt{2016A&A...586L...4D}).
Interactions due to the two spin-1/2 $^1$H atoms (with total $I = 1$ for ortho spin symmetry) are small and usually ignored.
As a result, five quantum numbers and the inversion symmetry ($J^{\pm}_{K_a,K_c},F_1, F$) are used to describe every hyperfine energy level. 
%The fitting routine estimates the total optical depth $\tau_t$, the FWHM linewidth $\Delta v$, and the velocity centroid $v_{\rm{LSR}}$. 
The peak optical depth in a hyperfine transition, $\tau_i$ is related to the total optical depth reported from the CLASS fit by $\tau_i = R_i \tau$ where $R_i$ is the normalized ($\sum_i R_i = 1$) relative hyperfine strength.
The relative hyperfine strengths are proportional to the square of the electric dipole matrix elements of the hyperfine transition summed over the degenerate magnetic states of the transition and are calculated from 6-j symbols derived using irreducible tensor methods and the Wigner-Eckert Theorem (see chapter 15 of \citealt{gordy1984microwave}) 
\begin{eqnarray}
R_i & = & \frac{(2F_1^\prime+1)(2F_1+1)
(2F^\prime +1)(2F +1)}{(2I_{\rm{N}} + 1)(2I_{\rm{D}} + 1)} \nonumber \\
& \times & \left\{ \begin{array}{ccc} 
1 & F_1^\prime & J^\prime \\ 
1   & J          & F_1 
\end{array}
\right\} ^2 
\left\{ \begin{array}{ccc} 
1 & F^\prime & F_1^\prime \\ 
1   & F_1      & F
\end{array}
\right\} ^2 \; ,
\end{eqnarray} 
where the prime superscripts indicate the quantum numbers of the upper energy level.  
The spontaneous emission coefficient is 
\begin{equation}%
    A_i = \frac{64\pi^4\nu_i^3}{3hc^3} \frac{S \mu^2}{2F^{\prime} + 1} (2I_{\rm{N}} + 1)(2I_{\rm{D}} + 1) \, R_i  \;,
\end{equation} 
where S $\approx 1.5$ is the unsplit asymmetric top line strength for our transition ($J^{\pm}_{K_a,K_c} = $ $1^+_{1,1}$ $-$ $1^-_{0,1}$) and $\mu = 1.463 \; \rm{D} = 1.463 \times 10^{-18}$ esu cm (\citealt{1982JMoSp..93...83C}) is the component of the electric dipole moment (along the c-axis) responsible for the transition. 
%The factor of 9 in both equations is from $(2I_{\rm{N}} + 1)(2I_{\rm{D}} + 1) = 9$.
The strongest hyperfine transition ($J^{\pm}_{K_a,K_c},F_1, F = $ $1^+_{1,1}$, 2, 3 $-$ $1^-_{0,1}$, 2, 3) has $R_i = 14/81$ and $A_i = 5.23 \times 10^{-6}$ s$^{-1}$ \citep{2016A&A...586L...4D}.
The hyperfine transitions used in our spectral fitting of the $J^{\pm}_{K_a,K_c} = $ $1^+_{1,1}$ $-$ $1^-_{0,1}$ transition of o-NH$_2$D are given in Table \ref{tab:HyperLevels}.

\begin{table*}
\centering
\caption{Hyperfine transitions of o-NH$_2$D $1_{11}^{+} \rightarrow 1_{01}^{-}$.  The velocity offsets, $\delta v_i$, are calculated from the frequencies in \citealt{2016A&A...586L...4D} with respect to a central frequency of $85926.278$ MHz.}
\label{tab:HyperLevels}
\begin{tabular}{ccrr}
\hline
F$_1^{\prime}$ F$^{\prime}$ $-$ F$_1$ F & $\nu_i$ & $\delta v_i$ & $R_i$ \\ 
& (MHz)             
& (km s$^{-1}$)  
\\
     \hline\hline
0	1	$-$	1	0	&	85924.692	&	5.5334	&		$	16	/	1296$	\\
0	1	$-$	1	2	&	85924.750	&	5.3311	&		$	80	/	1296$	\\
0	1	$-$	1	1	&	85924.782	&	5.2194	&		$	48	/	1296$	\\
2	1	$-$	1	0	&	85925.643	&	2.2155	&		$	20	/	1296$	\\
2	2	$-$	1	2	&	85925.662	&	2.1492	&		$	15	/	1296$	\\
2	3	$-$	1	2	&	85925.688	&	2.0585	&		$	84	/	1296$	\\
2	2	$-$	1	1	&	85925.694	&	2.0375	&		$	45	/	1296$	\\
2	1	$-$	1	2	&	85925.701	&	2.0131	&		$	1	/	1296$	\\
2	1	$-$	1	1	&	85925.733	&	1.9015	&		$	15	/	1296$	\\
2	2	$-$	2	2	&	85926.187	&	0.3175	&		$	125	/	1296$	\\
2	3	$-$	2	2	&	85926.213	&	0.2268	&		$	28	/	1296$	\\
2	1	$-$	2	2	&	85926.226	&	0.1814	&		$	27	/	1296$	\\
1	1	$-$	1	0	&	85926.243	&	0.1221	&		$	12	/	1296$	\\
2	2	$-$	2	3	&	85926.245	&	0.1151	&		$	28	/	1296$	\\
2	3	$-$	2	3	&	85926.271	&	0.0244	&		$	224	/	1296$	\\
1	2	$-$	1	2	&	85926.282	&	-0.0140	&		$	45	/	1296$	\\
2	2	$-$	2	1	&	85926.284	&	-0.0209	&		$	27	/	1296$	\\
1	0	$-$	1	1	&	85926.289	&	-0.0384	&		$	12	/	1296$	\\
1	1	$-$	1	2	&	85926.301	&	-0.0802	&		$	15	/	1296$	\\
1	2	$-$	1	1	&	85926.315	&	-0.1291	&		$	15	/	1296$	\\
2	1	$-$	2	1	&	85926.323	&	-0.1570	&		$	81	/	1296$	\\
1	1	$-$	1	1	&	85926.333	&	-0.1919	&		$	9	/	1296$	\\
1	2	$-$	2	2	&	85926.807	&	-1.8456	&		$	15	/	1296$	\\
1	1	$-$	2	2	&	85926.826	&	-1.9119	&		$	45	/	1296$	\\
1	2	$-$	2	3	&	85926.865	&	-2.0480	&		$	84	/	1296$	\\
1	0	$-$	2	1	&	85926.878	&	-2.0934	&		$	20	/	1296$	\\
1	2	$-$	2	1	&	85926.904	&	-2.1841	&		$	1	/	1296$	\\
1	1	$-$	2	1	&	85926.923	&	-2.2504	&		$	15	/	1296$	\\
1	0	$-$	0	1	&	85927.695	&	-4.9438	&		$	16	/	1296$	\\
1	2	$-$	0	1	&	85927.721	&	-5.0345	&		$	80	/	1296$	\\
1	1	$-$	0	1	&	85927.740	&	-5.1008	&		$	48	/	1296$	\\
\end{tabular}
\end{table*}

\section{Estimating Outer Density from Bonnor-Ebert Models}
\label{section:Appendix}

The density profile of an isothermal Bonnor-Ebert sphere at temperature $T$ is found by solving the differential equation
\begin{equation}
 \frac{1}{\xi^2} \frac{\rm{d}}{\rm{d}\xi} \left( \xi^2 \frac{\rm{d} \psi}{\rm{d} \xi} \right) = e^{- \psi(\xi)} \; ,
\end{equation}
where 
$\xi = r \left( \frac{4 \pi G \mu_p^2 m_{\rm{H}}^2 n_c}{k T} \right) ^{1/2}$ and $\psi(\xi) = - \ln \left( \frac{n(\xi)}{n_c} \right)$ are dimensionless variables related to the distance from the center, $r$, the density at each radius, and the central density, $n_c$.
The solution to the differential equation numerically calculates $e^{-\psi} = n(\xi)/n_c$ and $d\psi/d\xi$ for each $\xi$ \citep{1949ApJ...109..551C}. 
The mass of an isothermal Bonnor-Ebert sphere (in units of grams) is given by
\begin{equation}
M(\xi_{max}) = \frac{(kT/G)^{\frac{3}{2}}}{2\mu_p^2 m_{\rm{H}}^2 \sqrt{\pi n_c}}
                   \xi_{max}^2 \frac{d \psi(\xi_{max})}{d\xi} \; ,
\end{equation}
(see \citealt{2017stfo.book.....K} solutions to problem set 2 for a derivation).
$\xi_{max}$ is the value of the dimensionless radius at the boundary of the Bonnor-Ebert sphere.
The average density (cm$^{-3}$) may then be calculated from
\begin{equation}
    \bar{n} = \frac{3 M(\xi_{max})}{4 \pi \mu_p m_{\rm{H}} R^3} = \frac{3 \frac{(kT/G)^{\frac{3}{2}}}{2\mu_p^2 m_{\rm{H}}^2 \sqrt{\pi n_c}}
                   \xi_{max}^2 \frac{d \psi(\xi_{max})}{d\xi}}{4 \pi \mu_p m_{\rm{H}}  \left( \frac{kT/G}{4 \pi \mu_p^2 m_{\rm{H}}^2 n_c} \right)^{3/2} \xi_{max}^3 } = \frac{3 n_c \frac{d \psi(\xi_{max})}{d\xi}}{\xi_{max}} \; .
\end{equation}
Substituting for the central density, $n_c = n_o e^{\psi(\xi_{max})}$, from the definition of $\psi$, we then find that the ratio of the outer density to the  average density is given by
\begin{equation}
\frac{n_o}{\bar{n}} = \frac{\xi_{max} e^{-\psi(\xi_{max})}}{3 \frac{d\psi(\xi_{max})}{d\xi}}  \; .
\end{equation}
Figure \ref{fig:BES} plots $n_o/\bar{n}$ as a function of $\xi_{max}/\xi_{1/2}$ where $\xi_{1/2} = 2.274$ is the dimensionless radius at which the Bonnor-Ebert sphere has a density that is half the central density ($n(\xi_{1/2}) = n_c/2$).
A Bonnor-Ebert sphere is stable if $\bar{n}/n_o \leq 2.46247$ \citep{2001A&A...375.1091L}. 

%\section{Tables}

%%%%%%%%%%%%%%%%%%%%%%%%%%%%%%%%%%%%%%%%%%%%%%%%%%

% Don't change these lines
\bsp	% typesetting comment
\label{lastpage}
\end{document}